\documentclass[a4paper,11pt]{article}
\usepackage{jheppub}

\usepackage[utf8]{inputenc}
\usepackage[english]{babel}
\usepackage[T1]{fontenc}
\usepackage{amsmath,amssymb,graphicx, latexsym, theorem, enumerate,bbm}
\usepackage{booktabs, multirow, nicefrac, float, xspace, comment, array}
\usepackage[dvipsnames]{xcolor}
\usepackage{algorithm}
\usepackage{algpseudocode}
\usepackage[normalem]{ulem}
\usepackage{makecell}
\usepackage{pict2e}
\usepackage{shuffle}
\usepackage{subfigure}
\usepackage{feynmp}
\usepackage{feynmp-auto}
\usepackage{tikz,tikz-feynman}
\usetikzlibrary{arrows,decorations.markings,shapes.misc,math}

\allowdisplaybreaks[4]
\definecolor{darkpurple}{rgb}{0.5, 0.2, 0.8}
\definecolor{darkgreen}{rgb}{0.0, 0.4, 0.0}
\definecolor{darkyellow}{rgb}{0.5, 0.5, 0.0}
\definecolor{darkblue}{rgb}{0.0, 0.0, 0.8}
\definecolor{darkred}{rgb}{0.5, 0.0, 0.0}

\newcommand{\Reff}{R_\text{eff}}
\newcommand{\eps}{\varepsilon}
\newcommand{\zb}{\bar{z}}

\hypersetup{
    linktocpage,
     colorlinks,
     citecolor=darkgreen,
     linkcolor= darkgreen,
     urlcolor=darkgreen
}

\newcommand{\Li}{\text{Li}}
\newcommand{\cS}{ {\mathcal S} }
\newcommand{\cB}{ {\mathcal B} }
\newcommand{\bx}{ {\bf x} }
\newcommand{\dmin}{d_{\text{min}}}

\usepackage{ifpdf} 

\title{\boldmath 
Analytic Regression of Feynman Integrals from High-Precision
Numerical Sampling}

\author[a]{Oscar Barrera,}
\author[a,b]{Aur\'elien Dersy,}
\author[a]{Rabia Husain,}
\author[a,b]{Matthew D. Schwartz,}
\author[a]{and Xiaoyuan Zhang}
\affiliation[a]{Department of Physics, Harvard University, 02138 Cambridge, MA, USA}
\affiliation[b]{The NSF AI Institute for Artificial Intelligence and Fundamental Interactions}
\emailAdd{oscarbarrera@g.harvard.edu}
\emailAdd{adersy@g.harvard.edu}
\emailAdd{rhusain@g.harvard.edu}
\emailAdd{schwartz@g.harvard.edu}
\emailAdd{xiaoyuanzhang@g.harvard.edu}

\abstract{
In mathematics or theoretical physics one is often interested in obtaining an exact analytic description of some data which can be produced, in principle, to arbitrary accuracy. For example, one might like to know the exact analytical form of a definite integral. Such problems are not well-suited to numerical symbolic regression, since typical numerical methods  lead only to approximations. However, if one has some sense of the function space in which the analytic result should lie, it is possible to deduce the exact answer by judiciously sampling the data at a sufficient number of points with sufficient precision. We demonstrate how this can be done for the computation of Feynman integrals. We show that by combining high-precision numerical integration with analytic knowledge of the function space one can often deduce the exact answer using lattice reduction. A number of examples are given as well as an exploration of the trade-offs between number of datapoints, number of functional predicates, precision of the data, and compute. This method provides a bottom-up approach that neatly complements the top-down Landau-bootstrap approach of trying to constrain the exact answer using the analytic structure alone. Although we focus on the application to Feynman integrals, the techniques presented here are more general and could apply to a wide range of problems where an exact answer is needed and the function space is sufficiently well understood.
}

\begin{document}
\unitlength = 0.4mm

\newcommand{\triangleoneloop}[3]{
\mbox{\parbox{3cm}{\hspace{0.25cm}
\begin{picture}(2.5,1.4)
\put(0.3,0.7){\vector(1,0){0.1}}
\put(1.9,0.2){\vector(1,0){0.1}}
\put(1.9,1.2){\vector(1,0){0.1}}
\put(0.5,0.7){\line(2,1){1}}
\put(0.5,0.7){\line(2,-1){1}}
\put(1.5,1.2){\line(0,-1){1}}
\linethickness{0.35mm}
\put(1.5,1.2){\line(1,0){0.5}}
\put(0,0.7){\line(1,0){0.5}}
\put(1.5,0.2){\line(1,0){0.5}}
\thinlines
\put(0.25,0.9){\makebox(0,0)[b]{$#1$}}
\put(2.05,1.2){\makebox(0,0)[l]{$#2$}}
\put(2.05,0.2){\makebox(0,0)[l]{$#3$}}
\end{picture}
}}
\hfill}

\newcommand{\triangletwosfourd}[4]{
\mbox{\parbox{3cm}{\hspace{0.25cm}
\begin{picture}(2.5,1.4)
\put(0.3,0.7){\vector(1,0){0.1}}
\put(1.7,0.2){\vector(1,0){0.1}}
\put(1.7,1.2){\vector(1,0){0.1}}
\put(0,0.7){\line(1,0){0.5}}
\linethickness{0.35mm}
\put(1,1.2){\line(1,0){1}}
\put(1,0.2){\line(1,0){1}}
\thinlines
\put(1,1.2){\line(0,-1){1}}
\put(1.0,0.7){\circle*{0.2}}
\put(1,0.7){\circle{1}}
\put(0.25,0.9){\makebox(0,0)[b]{$#1$}}
\put(2.05,1.2){\makebox(0,0)[l]{$#2$}}
\put(2.05,0.2){\makebox(0,0)[l]{$#3$}}
\put(1.1,-0.5){\makebox(0,0)[b]{$#4$}}
\end{picture}
}}
\hfill}

\newcommand{\triangletwosfived}[4]{
\mbox{\parbox{3cm}{\hspace{0.25cm}
\begin{picture}(2.5,1.4)
\put(0.3,0.7){\vector(1,0){0.1}}
\put(1.9,0.2){\vector(1,0){0.1}}
\put(1.9,1.2){\vector(1,0){0.1}}
\put(0.5,0.7){\line(2,1){1}}
\put(0.5,0.7){\line(2,-1){1}}
\put(1.5,1.2){\line(0,-1){1}}
\put(0,0.7){\line(1,0){0.5}}
\linethickness{0.35mm}
\put(1.5,1.2){\line(1,0){0.5}}
\put(1.5,0.2){\line(1,0){0.5}}
\thinlines
\put(1.5,0.2){\line(-1,2){0.4}}
\put(0.25,0.9){\makebox(0,0)[b]{$#1$}}
\put(2.05,1.2){\makebox(0,0)[l]{$#2$}}
\put(2.05,0.2){\makebox(0,0)[l]{$#3$}}
\put(1.4,-0.5){\makebox(0,0)[b]{$#4$}}
\end{picture}
}}
\hfill}

\newcommand{\boxthreesfived}[5]{
\mbox{\parbox{3.5cm}{\hspace{0.25cm}
\begin{picture}(3,1.4)
\put(0.8,0.2){\vector(1,0){0.1}}
\put(2.3,0.2){\vector(1,0){0.1}}
\put(2.3,1.2){\vector(1,0){0.1}}
\put(0.8,1.2){\vector(1,0){0.1}}
\put(1.5,1.2){\oval(1,1)[b]}
\put(0.5,0.2){\line(1,0){1.5}}
\put(0.5,1.2){\line(1,0){1.5}}
\linethickness{0.35mm}
\put(2.0,0.2){\line(1,0){0.5}}
\put(2.0,1.2){\line(1,0){0.5}}
\thinlines
\put(1,0.2){\line(0,1){1}}
\put(2,0.2){\line(0,1){1}}
\put(0.45,1.2){\makebox(0,0)[r]{$#1$}}
\put(0.45,0.2){\makebox(0,0)[r]{$#2$}}
\put(2.55,1.2){\makebox(0,0)[l]{$#3$}}
\put(2.55,0.2){\makebox(0,0)[l]{$#4$}}
\put(1.5,-0.5){\makebox(0,0)[b]{$#5$}}
\end{picture}
}} 
\hfill}

\newcommand{\boxthreessevend}[5]{
\mbox{\parbox{3.5cm}{\hspace{0.25cm}
\begin{picture}(3,1.4)
\put(0.8,0.2){\vector(1,0){0.1}}
\put(2.3,0.2){\vector(1,0){0.1}}
\put(2.3,1.2){\vector(1,0){0.1}}
\put(0.8,1.2){\vector(1,0){0.1}}
\put(0.5,0.2){\line(1,0){1.5}}
\put(0.5,1.2){\line(1,0){1.5}}
\linethickness{0.35mm}
\put(2.0,0.2){\line(1,0){0.5}}
\put(2.0,1.2){\line(1,0){0.5}}
\thinlines
\put(1,0.2){\line(0,1){1}}
\put(1.5,0.2){\line(0,1){1}}
\put(2,0.2){\line(0,1){1}}
\put(0.45,1.2){\makebox(0,0)[r]{$#1$}}
\put(0.45,0.2){\makebox(0,0)[r]{$#2$}}
\put(2.55,1.2){\makebox(0,0)[l]{$#3$}}
\put(2.55,0.2){\makebox(0,0)[l]{$#4$}}
\put(1.5,-0.5){\makebox(0,0)[b]{$#5$}}
\end{picture}
}}
\hfill}

\newcommand{\outermassdbox}[5]{
\mbox{\parbox{5cm}{\hspace{0.5cm}
\begin{picture}(3,1.8)
\put(0.59,-0.1){\vector(4,3){0.1}}
\put(3.9,-0.1){\vector(-4,3){0.1}}
\put(3.9,2.0){\vector(-4,-3){0.1}}
\put(0.59,2.0){\vector(4,-3){0.1}}
\put(0.255,2.25){\line(4,-3){0.75}}
\put(3.5,0.2){\line(4,-3){0.75}}
\put(3.5,1.7){\line(4,3){0.75}}
\put(0.255,-0.35){\line(4,3){0.75}}
\put(2.25,0.2){\line(0,1){1.5}}
\linethickness{0.5mm}
\put(3.5,0.2){\line(0,1){1.5}}
\put(1,0.2){\line(0,1){1.5}}
\put(1,0.2){\line(1,0){2.5}}
\put(1.0,1.7){\line(1,0){2.5}}
\put(0.8,1.7){\makebox(0,0)[r]{$#1$}}
\put(0.8,0.2){\makebox(0,0)[r]{$#2$}}
\put(3.75,1.7){\makebox(0,0)[l]{$#3$}}
\put(3.75,0.2){\makebox(0,0)[l]{$#4$}}
\put(2,-0.5){\makebox(0,0)[b]{$#5$}}
\end{picture}
}}
\hfill}

\newcommand{\triangleonel}[3]{
\mbox{\parbox{4cm}{\hspace{0.35cm}
\begin{picture}(3,1.4)
\put(0.3,0.7){\vector(1,0){0.1}}
\put(2.25,-0.175){\vector(-2,1){0.1}}
\put(2.25,1.575){\vector(-2,-1){0.1}}
\put(0.5,0.7){\line(2,1){1.5}}
\put(0.5,0.7){\line(2,-1){1.5}}
\put(2,1.45){\line(0,-1){1.5}}
\linethickness{0.35mm}
\put(2,1.45){\line(2,1){0.5}}
\put(0,0.7){\line(1,0){0.5}}
\put(2,-0.05){\line(2,-1){0.5}}
\thinlines
\put(0.25,0.9){\makebox(0,0)[b]{$#1$}}
\put(2.2,1.3){\makebox(0,0)[l]{$#2$}}
\put(2.2,0.1){\makebox(0,0)[l]{$#3$}}
\end{picture}
}}
\hfill}

\newcommand{\triangletwol}[3]{
\mbox{\parbox{4cm}{\hspace{0.35cm}
\begin{picture}(3,1.4)
\put(0.3,0.7){\vector(1,0){0.1}}
\put(2.25,-0.175){\vector(-2,1){0.1}}
\put(2.25,1.575){\vector(-2,-1){0.1}}
\put(0.5,0.7){\line(2,1){1.5}}
\put(0.5,0.7){\line(2,-1){1.5}}
\put(2,1.45){\line(0,-1){1.5}}
\put(1.5,1.2){\line(0,-1){1.0}}
\linethickness{0.35mm}
\put(2,1.45){\line(2,1){0.5}}
\put(0,0.7){\line(1,0){0.5}}
\put(2,-0.05){\line(2,-1){0.5}}
\thinlines
\put(0.25,0.9){\makebox(0,0)[b]{$#1$}}
\put(2.2,1.3){\makebox(0,0)[l]{$#2$}}
\put(2.2,0.1){\makebox(0,0)[l]{$#3$}}
\end{picture}
}}
\hfill}

\newcommand{\trianglethreel}[3]{
\mbox{\parbox{4cm}{\hspace{0.35cm}
\begin{picture}(3,1.4)
\put(0.3,0.7){\vector(1,0){0.1}}
\put(2.25,-0.175){\vector(-2,1){0.1}}
\put(2.25,1.575){\vector(-2,-1){0.1}}
\put(0.5,0.7){\line(2,1){1.5}}
\put(0.5,0.7){\line(2,-1){1.5}}
\put(2,1.45){\line(0,-1){1.5}}
\put(1.7,1.3){\line(0,-1){1.2}}
\put(1.3,1.1){\line(0,-1){0.8}}
\linethickness{0.35mm}
\put(2,1.45){\line(2,1){0.5}}
\put(0,0.7){\line(1,0){0.5}}
\put(2,-0.05){\line(2,-1){0.5}}
\thinlines
\put(0.25,0.9){\makebox(0,0)[b]{$#1$}}
\put(2.2,1.3){\makebox(0,0)[l]{$#2$}}
\put(2.2,0.1){\makebox(0,0)[l]{$#3$}}
\end{picture}
}}
\hfill}

\maketitle
\flushbottom

\newpage

\section{Introduction}
Symbolic regression aims to provide a minimal, but usually approximate, functional description of possibly noisy data. Such descriptions can be enormously helpful in biology or astronomy, for example. In mathematics or theoretical physics, on the other hand, one is often interested in an exact functional description of data which can be obtained in principle to arbitrary precision.  That is, one would like the exact analytic function with no approximate real number coefficients scattered within the analytic functional predicates. If one has in addition some knowledge of the possible form of the function, then one might hope to find the answer through analytic regression. In this paper we will show that, for the particular case of the computation of Feynman integrals where the space of functions is known one can often determine the answer exactly.

The problem we are concerned with is the decomposition of a multivariate function $f(\bx)$ in terms of $n$ known basis functions $\cB_i(\bx)$
\begin{equation}\label{eq:problem_statement}
f(\mathbf{x}_j) = \sum_{i=1}^n c_i \,\mathcal{B}_i(\mathbf{x}_j) \,, 
\end{equation}
where $\mathbf{x}_j$ are points in some number of dimensions and $c_i$ are rational numbers, some of which may be zero. If $f(\bx)$ and $\cB_i(\bx)$ can be evaluated exactly at $n$ points then we can simply solve for $c_i$ by inverting the matrix $M_{ij} = \cB_i(\bx_j)$. However, if $f(\bx)$ and $\cB_i(\bx)$ are only known numerically to a minimum precision of $d$ digits, then inverting $M_{ij}$ will not give the exact result and in general the $c_i$ produced will have much fewer than $d$ digits of precision. We want instead {\it exact} rational $c_i$ using numerical values of $f(\bx)$ and $\cB_i(\bx)$.

As a concrete application, consider the 1-loop triangle Feynman integral:
\begin{equation}
T_1(p_1^2,p_2^2,p_3^2)  =
\vcenter{\hbox{ 
\begin{tikzpicture}[scale=0.5, baseline={(current bounding box.center)}]
\begin{feynman}
    \vertex (v1) at (0, 0);
    \vertex (v2) at (2, 1);
    \vertex (v3) at (2,-1); 
    \vertex (p1) at (-1.5, 0);
    \vertex (p2) at (3.5, 1);
    \vertex (p3) at (3.5,-1); 
    \diagram* {
        (p1) -- [thick,, momentum'=\(p_1\)] (v1)
             --  (v2)
             --  (v3)
             --  (v1),
        (p2) -- [thick,  momentum'=\(p_2\)] (v2),
        (p3) -- [thick, momentum=\(p_3\)] (v3),
    };
\end{feynman}
\end{tikzpicture}
}}
\hspace{12pt}= \int\frac{d^4 k}{i \pi^2}\frac{1}{k^2+i\eps} \frac{1}{(p_2-k)^2 +i\eps} \frac{1}{(p_3+k)^2+i\eps}\,.
\end{equation}  
This integral is known exactly and most judiciously expressed in terms of the variables  $z$ and $\zb$ which solve $z \zb \ = p_2^2/p_1^2$ and $(1-z)(1-\zb) = p_3^2/p_1^2$. Explicitly,
\begin{equation}
   p_{1}^{2}  T_1(p_1^2,p_2^2,p_3^2) =
 2\frac{\Li_2(z)}{z-\zb} -2\frac{\Li_2(\zb)}{z-\zb}+\frac{\ln (z \zb) \ln \left(\frac{1-z}{1-\zb}\right)}{z-\zb}\,.
 \label{eq:zform}
\end{equation} 
In this context, what we would like to achieve is the regression of the exact coefficients $c_i = \{2,-2,1\}$  given the basis functions on the right as well as their numerical evaluation and the evaluation of $T_1$ at a sufficient number of phase space points with sufficient precision. This example is discussed in Section~\ref{sec:ladders}.

For this program to be successful we first need 1) a methodology to  determine the finite set of basis functions $\cB_i(\bx)$ and 2) a methodology that can evaluate them and the true function $f(\bx)$ to high precision in reasonable time. For the computation of Feynman diagrams, there has been extraordinary progress in both of these directions over the last two decades or so. The key to the first step is that the possible basis functions are strongly constrained by symmetries and the singularity structure of the Feynman integral. 
Singularities of Feynman rules are given by solutions to the Landau equations~\cite{Landau:1959fi}. Solving these equations was a key component of the analytic $S$-matrix program of mid 20th century~\cite{ELOP,Steinmann,Steinmann2,pham} and has received renewed attention in the 21st century
~\cite{Fevola:2023kaw,Hannesdottir:2021kpd,Hannesdottir:2022xki}.
Combined with improved knowledge of the types of functions that can arise from Feynman integrals~\cite{Broadhurst:1998rz,Bogner:2007mn,Bourjaily:2022bwx}
this has allowed for the possibility of determining a Feynman integral completely from its analytic and algebraic properties alone
\cite{Arkani-Hamed:2012zlh,Schlotterer:2012ny,Brown:2015fyf,Panzer:2016snt,Schnetz:2017bko,Caron-Huot:2019bsq,Gurdogan:2020ppd,Dixon:2021tdw,Dixon:2022xqh}. Following~\cite{Hannesdottir:2024hke} we refer to this program as the Landau bootstrap. 
There are now public software packages such as \textsc{SOFIA}~\cite{Correia:2025yao} that automate the determination of singularities and provide natural variables on which a Feynman integral should depend. 
The Landau bootstrap then proceeds by imposing algebraic and physical constraints of increasing sophistical complexity until the integral is completely fixed.  See Ref.~\cite{Hannesdottir:2024hke} for some examples. 

The second remarkable development concerns improvements in the ability to compute Feynman integrals numerically. Naively, a  finite $\ell$-loop integral in $d=4$ dimensions requires $4\ell$ integrations. One can reduce to an integral over $n-1$ Feynman parameters, where $n$ is the number of propagators, or $n-\ell$ Baikov variables. However, with any such multidimensional integral one can only expect a few digits of precision with reasonable computation cost. Moreover, most integrals of interest also have ultraviolet (UV) or infrared (IR) divergences. Modern techniques however can reduce the computation to numerically solving a one-dimensional differential equation. Consequently we now have access to automated tools such as \textsc{AMFlow}~\cite{Liu:2017jxz, Liu:2018dmc, Liu:2021wks,Liu:2022chg, Liu:2022mfb} which can efficiently provide the numerical evaluation of any term in the $\epsilon$-expansion of a Feynman integral to tens or hundreds of digits. 

With tools that can provide for us both sides of Eq.~\eqref{eq:problem_statement} in a semi-automated way, we turn to solving this equation for the rational coefficients.
There are many approaches one might take. The simplest, as mentioned, is to invert $M_{ij} = \cB_i(\bx_j)$ which gives real $c_i$ which then can be rounded into nearby rational numbers. This approach is problematic as error propagation does not scale well with the number $n$ of basis functions. Instead, we advocate an approach based on {\it lattice reduction}. A $d$-dimensional lattice can be defined by any set of $d$ independent basis vectors. Lattice reduction attempts to reduce a given set of basis vectors to a new set which minimizes some norm, such as the sum of the lengths of the basis vectors. There are many algorithms for doing so, such as Lenstra–Lenstra–Lovász (LLL)~\cite{Lenstra:1982} or its refinements (we briefly review these in Appendix~\ref{app:lll_details}). 

In fact, lattice reduction has been used for decades already in the computation of Feynman integrals to fit linear combinations of transcendental numbers. 
For example, suppose we have an integral like
\begin{equation}
f=   - \int _0^1 du \int _0^1 dv \frac{\log (1-u v)+v \log (1-u)}{u v} = \frac{\pi^2}{6} +\zeta_3 \,.
\label{fnums}
\end{equation}
In this case there is no freedom in choosing the phase space points -- the integral is just a number. However, if we know the possible transcendental numbers which can appear on the right-hand side, as we often do, their coefficients can still be fit using lattice reduction. 
To see how this works, suppose we have four digits of accuracy. Then $f=2.847$, $\pi^2 = 9.870$ and $\zeta_3 = 1.202$. Multiplying by $10^3$, putting these numbers into a matrix, and applying lattice reduction gives
\begin{equation}
  \left(\begin{array}{c}
    \vec{u}_1\\
    \vec{u}_2\\
    \vec{u}_3
  \end{array}\right) = \left(\begin{array}{cccc}
    2847 & 1 & 0 & 0 \\
    9870 &0 & 1 & 0\\
    1202 &0 & 0 & 1
  \end{array}\right) \xrightarrow[\text{lattice reduction}]{}
  \left(\begin{array}{c}
    \vec{v}_1\\
    \vec{v}_2\\
    \vec{v}_3
  \end{array}\right) = \left(\begin{array}{cccc}
    0 & - 6 & 1 & 6\\
    3 & - 11 & 5 & - 15 \\
    62 & 4 & - 2 & 7 
  \end{array}\right) \,.
  \label{eq:llexample}
\end{equation}
This means that the three 4-vectors on the right span the same lattice as the three 4-vectors on the left. In particular, $\vec{v}_1$ is a linear combination of the $\vec{u}_i$ with integer coefficients. Considering the $3 \times 3$ identity matrix at the end of the $\vec{u}$ matrix, we must then have
$\vec{v}_1 = - 6 \vec{u}_1 + \vec{u}_2 + 6 \vec{u}_3$. The 4th component of this equation is $0 = 10^3 \times (- 6 f + \pi^2 + 6 \zeta^3)$ so that $f = \pi^2/6 + \zeta_3$. 
For this problem of fitting pure numbers one can also use the PSLQ algorithm~\cite{Ferguson:1992,PSLQref,Bailey:1999nv}. 
Fitting linear combinations of transcendental numbers with lattice reduction or PSLQ has been extensively used in particle physics~\cite{Chachamis:2008fx,Lee:2010cga,Kelley:2011ng,Laporta:2017okg,Chicherin:2017dob, Lee:2018ojn,Acres:2021sss, Agarwal:2021zft, Canko:2024ara}.
Our work is inspired by these successes: we want to generalize from fitting numbers this way to fitting functions.

A drawback of lattice reduction is that in general one needs high numerical accuracy. In typical applications, one often needs 100 digits or more of numerical precision in the integrals to fit a handful of transcendental numbers. If we want to fit 1000 functions we might then need 10,000 digits of precision! A key insight of this paper is that one can trade higher precision for sampling over more points. Naively, there is the same information in $d$ digits for a single number as $\frac{d}{p}$ digits in $p$ numbers, so we can evaluate the integral over 100 points with 100 digits instead of 1 point with 10,000 digits. Section~\ref{sec:p_d_scaling} provides more details of the scaling and tradeoffs.  

\begin{figure}
    \centering
    \includegraphics[width=\linewidth]{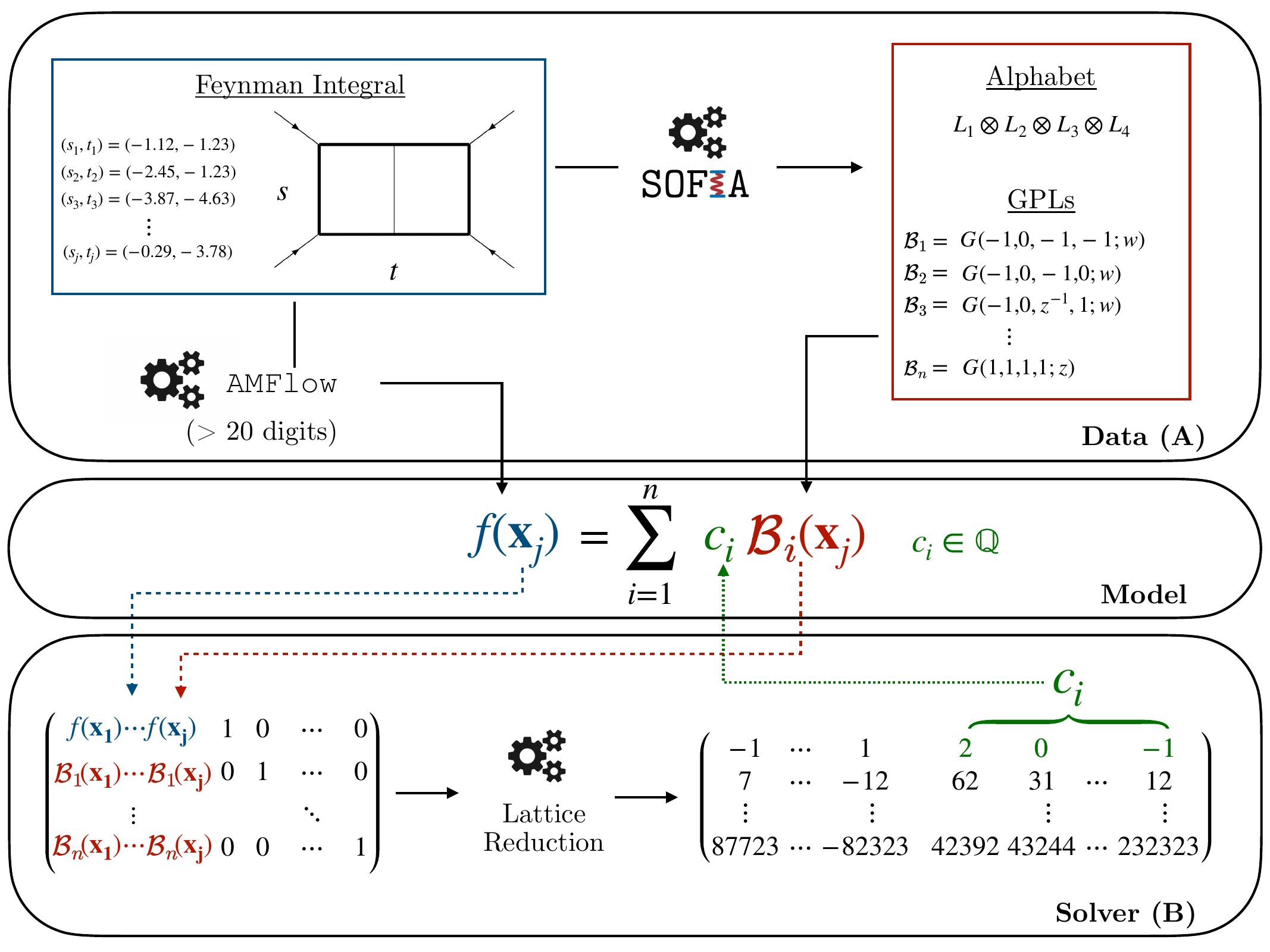}
    \caption{To recover the analytic expression associated with a Feynman integral, we proceed in two steps. First (module A) we numerically evaluate the integral at high precision with \textsc{AMFlow}~\cite{Liu:2017jxz, Liu:2018dmc, Liu:2021wks,Liu:2022chg, Liu:2022mfb} for multiple kinematic points and construct the appropriate basis of functions from the singularity analysis in \textsc{SOFIA}~\cite{Correia:2025yao}. Then (module B) we perform the linear fit using lattice reduction, analytically recovering the appropriate rational coefficients.}
    \label{fig:flowchart}
\end{figure}

Fig.~\ref{fig:flowchart} gives an overview of our vision of analytic regression. 
There are two main modules: module A produces the input (previous work) and module B performs the analytic regression (current work).
The inputs needed (module A) are the evaluation of the function $f({\bf{x}})$ and the basis functions $\cB_i({\bf{x}})$  at various points to some desired precision. 
For Feynman integrals, $f({\bf{x}})$ can be computed using automated tools like \textsc{AMFlow} and $\cB_i({\bf{x}})$ constructed with other tools like \textsc{SOFIA}.  Once the numbers are computed, they can then be input into the analytic regression algorithm indicated as module B on the bottom of the figure.
Although the focus of this paper will be on the analytic regression of Feynman integrals, the methodology could be applied to many problems in mathematics or other areas of physics where an exact functional form is needed and could be extracted from numerical evaluations. 
For an alternative application, one simply swaps out module A leaving module B untouched.

The outline of this paper is as follows. We begin in Section~\ref {sec:fit_Feyn_int} with a discussion of module A for the production of Feynman integrals. We establish our notation and discuss how one can now compute a Feynman integral numerically. We also discuss how one can use public tools to construct a basis in a semi-automated way. Section~\ref{sec:lll_algo} discusses the matrix inversion and lattice reduction approaches to module B. Section~\ref{sec:simple_integrals} gives a number of end-to-end worked examples up to 3-loops. After the examples, in Section~\ref{sec:lll_limitations}, we discuss some trade-offs and scaling behavior. Conclusions are in Section~\ref{sec:conclusion}.

\section{Computing $f(\bx)$ and $\cB_i(\bx)$ from Feynman integrals}\label{sec:fit_Feyn_int}
As explained in the introduction, we are interested in solving Eq.~\eqref{eq:problem_statement} where the function to fit $f(\bx)$ is a Feynman integral. In this section we discuss the module A in detail, namely how these integrals can be computed and an appropriate basis determined and evaluated numerically as well. 

\subsection{Numerical evaluation of $f(\bx)$}
Feynman integrals are one of the most important objects appearing in quantum field theories and particle physics. An $L$-loop $d$-dimensional scalar Feynman integral with loop momenta $\ell_i$ and external momenta $p_k$ can be written as
\begin{equation}
\label{eq:def_loop}
f(p_k) = 
e^{L\epsilon \gamma_E}\int\prod_{i=1}^{L}\frac{d^d \ell_i}{i\pi^{d/2}}\frac{\mathcal{P}(\ell_1\cdot \ell_2,\cdots;\ell_1\cdot p_1,\cdots)}{D_1^{\alpha_1}\, D_2^{\alpha_2}\,\cdots D_n^{\alpha_n}}\,,
\end{equation}
where $D_j=q_j^2-m_j^2+i\epsilon$ represents the propagator in a Feynman diagram and $\alpha_j$ is its associated power. The propagators depend on $q_j$, which are given as linear combinations of the loop and external momenta. The numerator $\mathcal{P}(\ell_1\cdot \ell_2,\cdots;\ell_1\cdot p_1,\cdots)$ is a polynomial in the external and loop momenta. The result is a scalar function of Lorentz-invariant inner products of the various external momenta $\bx =\{p_i\cdot p_j\}$ and the masses $m_j$.

Before being able to evaluate a Feynman integral numerically, one must first regulate its possible ultraviolet (UV)  and infrared (IR) divergences. A standard way of dealing with such divergences is to introduce a dimensional regularization parameter $d=4-2\epsilon$, and perform the integrals in $\epsilon\to 0$ expansion. Naively, the $\epsilon$ expansion can only be performed after the integration is finished. However, various techniques have been developed to decouple different divergences such that it becomes safe to first expand the integrand in $\epsilon$. This is usually referred to as sector decomposition \cite{Binoth:2000ps, Binoth:2003ak, Binoth:2004jv, Heinrich:2008si}, and it is implemented in packages like \textsc{Fiesta}~\cite{Smirnov:2008py,Smirnov:2009pb,Smirnov:2013eza,Smirnov:2015mct,Smirnov:2021rhf} and \textsc{PySecDec}~\cite{Borowka:2012yc, Borowka:2015mxa, Borowka:2017idc, Borowka:2018goh, Heinrich:2021dbf, Heinrich:2023til}.

One of the big breakthroughs for calculating Feynman integrals came with the Integration-by-Parts (IBPs) technique~\cite{Tkachov:1981wb,Chetyrkin:1981qh}, which, given a topology, allows for expressing any Feynman integrals living in this topology in terms of a linear combination of an integral basis. This integral basis is referred to as the master integral basis. The IBP algorithms have been implemented in different public packages, for example, \textsc{LiteRed}~\cite{Lee:2012cn,Lee:2013mka}, \textsc{Fire}~\cite{Smirnov:2008iw,Smirnov:2014hma,Smirnov:2019qkx, Smirnov:2023yhb}, \textsc{Kira}~\cite{Maierhofer:2017gsa,Klappert:2020nbg,Lange:2025fba}, \textsc{Blade}~\cite{Liu:2018dmc, Guan:2019bcx, Guan:2024byi}, and \textsc{AmpRed}~\cite{Chen:2024xwt}. Once the master integrals are evaluated, all integrals within the same topology are available. To compute the master integrals, the natural approach is to form a closed differential equation (DE)~\cite{Kotikov:1990kg,Kotikov:1991pm,Bern:1993kr,Remiddi:1997ny, Henn:2013pwa} that is satisfied by the masters and express the analytic expressions in terms of classical/generalized/elliptical polylogarithms~\cite{Goncharov:1998kja, Goncharov:2001iea, Goncharov:2010jf, Duhr:2011zq, Duhr:2012fh}. However, solving the DE is non-trivial since (1) one needs to disentangle the $\epsilon$ dependence from kinematic dependence by searching for the $\epsilon$-factorized form; (2) finding a good boundary condition is not straightforward. 

Although the analytic evaluation of Feynman integrals still remains challenging, high-precision numerical evaluation has become available in recent years.
For instance, the public program \textsc{AMFlow}~\cite{Liu:2017jxz, Liu:2018dmc, Liu:2021wks,Liu:2022chg, Liu:2022mfb} pushes the numerical precision much higher than \textsc{Fiesta} or \textsc{PySecDec} using the IBP+DE technique. There, an auxiliary mass parameter $\eta$ is introduced in some of the propagators and DEs for $\eta$ are derived through IBP. Solving the DEs with boundary conditions at $\eta\to\infty$ gives high-precision numerical results for the associated Feynman integrals. This technique can be applied to compute either boundary conditions for DE or values of the Feynman integral itself. In the context of collider physics, to reduce computation cost, one can use \textsc{AMFlow} to get the value for one phase-space point and numerically solve the DEs for kinematic variables to obtain others. However, for convenience, we will compute all phase-space points in \textsc{AMFlow} directly in the following examples.

\subsection{Basis functions $\mathcal{B}_i(\mathbf{x})$\label{sec:basis_set}}
Recovering the result of a Feynman integral without actually computing it has been an active area of research in recent decades and can be considered part of a broader $S$-matrix bootstrap program. The type of bootstrap relevant for the current work involves first enumerating a finite set of possible transcendental functions allowed by the singularities of an amplitude or integral,
and then refining the allowed set by imposing physical constraints or consistency conditions. This can be done at the amplitude level~\cite{Dixon:2016nkn,Caron-Huot:2016owq,Caron-Huot:2019vjl,Dixon:2020bbt,Dixon:2022rse,Basso:2024hlx,Cai:2024znx,Cai:2025atc,Guo:2021bym}, where all relevant Feynman diagrams are summed over. At the amplitude level, one has access to constraints such as supersymmetry, collinear limits and gauge invariance which are violated by individual diagrams. For example, in Refs.~\cite{Dixon:2020bbt, Dixon:2022rse}, an amplitude bootstrap program was developed to predict a three-point form factor up to 8 loops in planar $\mathcal{N}=4$ super Yang-Mills theory. One can also apply the bootstrap to individual Feynman integrals~\cite{Chicherin:2017dob,Caron-Huot:2018dsv,Henn:2018cdp,He:2021fwf,He:2021eec,Morales:2022csr}, rather than amplitudes. For Feynman integrals, one does not have so much access to physical constraints, like gauge-invariance, but one can directly apply techniques like Landau analysis to the integral to more directly access its analytic structure.
This approach has been called the Landau bootstrap in~\cite{Hannesdottir:2024hke} and has been used in various forms for decades.

To fit a Feynman integral, we will leverage the fact that its analytical expression is heavily constrained by its singularity structure. In particular, for a wide class of integrals, it is possible to systematically construct a complete set of basis functions. For all the Feynman integrals we consider here, the result can be expressed in terms of Generalized (or Goncharov or multiple) Polylogarithms~\cite{Goncharov:1998kja, Goncharov:2010jf, Duhr:2012fh} (GPLs). Other functions, such as elliptical polylogarithms, may be required to describe other classes of Feynman integrals \cite{Georgoudis:2015hca, Bourjaily:2018ycu, Duhr:2022dxb, Bourjaily:2022bwx}. The approach we develop in this paper would apply perfectly well to these other function classes. However, since less is understood about these other classes, we focus on multiple-polylogarithmic amplitudes for simplicity.

To construct the basis of appropriate GPLs, we will need to first reconstruct the relevant alphabet, a set of algebraic functions $\{W_i(\mathbf{x})\}$ that can appear in the iterated integrals and in the arguments of the GPLs.  Often the alphabet can be determined completely in an automated way, for example by using principal Landau determinants~\cite{Fevola:2023fzn} as implemented in the \textsc{SOFIA} package \cite{Correia:2025yao}. With the alphabet in hand we then write down the set of integrable symbols\footnote{We provide a brief introduction to the symbol formalism in Appendix~\ref{sec:symbols} and describe how to obtain the subset of integrable symbols.} up to transcendental weight $2 L$, with $L$ the number of loops in the Feynman integral. When the alphabet is multilinear in the kinematic variables, going from an integrable symbol to a GPL is easily done with the \textit{FiberSymbol} routine in \textsc{PolyLogTools} \cite{Duhr:2019tlz}, which is based on the algorithm of Ref.~\cite{Anastasiou:2013srw}. We refer to the obtained GPL bases as $\{\tilde{\mathcal{B}}_i(\mathbf{x})\}$. These GPLs can then be numerically evaluated to arbitrary precision\footnote{If less than 16 digits of precision are required, we note that the \textsc{FastGPL} implementation~\cite{Wang:2021imw} provides a fast numerical evaluation. If more than 16 digits are required or if \textsc{FastGPL} does not guarantee the convergence of the series representation, we default to \textsc{GiNaC}.} using the \textsc{GiNaC} wrapper~\cite{Bauer:2002, Vollinga:2004sn}. If the alphabet is not multilinear in the kinematic variables and the GPLs are not easily obtainable, it is still possible to have a numerical evaluation for a particular symbol. Indeed, as we will describe in Section~\ref{sec:dbox}, one can assign an iterated integral to a particular symbol by specifying a choice of boundary conditions. These iterated integrals can then be done numerically and can serve as the set of relevant basis functions. 
Finally, we note that the basis $\{\tilde{\mathcal{B}}_i(\mathbf{x})\}$ will in general differ from $\{\mathcal{B}_i(\mathbf{x})\}$ in Eq.~\eqref{eq:problem_statement} by a rational prefactor. Assuming this prefactor is universal,\footnote{If there are more propagators than integrations then we can have multiple leading singularities and associated algebraic prefactors. In that case we would need to use the basis function set given by the outer product of $\{P_j(\mathbf{x})\} \times \{\mathcal{B}_i(\mathbf{x})\}$. In this work we only consider integrals with a single uniform prefactor.} we  have
\begin{equation}
\label{BBform}
    f(\mathbf{x}) = \sum_{i=1}^n c_i \,\mathcal{B}_i(\mathbf{x}) = P(\mathbf{x}) \sum_{i=1}^n c_i \,\tilde{\mathcal{B}}_i(\mathbf{x}) \,.
\end{equation}
For example, in Eq.~\eqref{eq:zform}, the prefactor for fitting $T_1$ is $P(\bx) = [p_1^2 (z-\zb) ]^{-1}$.
One can fix this algebraic prefactor by computing the leading singularity of the Feynman integral, utilizing the maximal cut \cite{Primo:2016ebd}. This calculation can also be done using the \textsc{SOFIA} package.

\section{Analytic regression of $c_i$}
\label{sec:lll_algo}
With a procedure to compute $f(\bx)$ and $\cB_i(\bx_j)$ to $d$ digits of precision at points $\bx_j$ we now turn to the analytic regression of the coefficients in Eq.~\eqref{eq:problem_statement}. We discuss the naive matrix inversion approach and then our main tool, lattice reduction. Further analysis of the lattice reduction approach is provided in Section~\ref{sec:lll_limitations}.

\subsection{Matrix inversion \label{sec:mi}}
The most basic way to determine $c_i$ is by inverting the matrix $M_{ij} = \cB_i(\bx_j)$
 to get $\tilde{c}_i =( M^{-1})_{ij} f(\bx_j)$.
 Since the matrix is real (or complex) the coefficients $\tilde{c}_i$ will be real (or complex) and then must be rounded and rationalized to get a candidate solution $c_i$.
 While this should work in principle, in practice it can be challenging to use.

The first problem is that matrix has no way to exploit the information that the $c_i$ are rational numbers typically of order 1. This problem is not just about efficiency but can lead the method to fail even with infinite precision if functions are proportional to one another, with the constant of proportionality being a transcendental number. This can be the case for Feynman integrals. For instance if the basis set contains
$\pi \ln x$ and $\frac{\pi^2}{6} \ln x$ (as do some examples from Section~\ref{sec:masters} below) and the expected coefficients are $\{0,-2\}$, matrix inversion could instead give back $\{-0.5236..., -1.0000\}\sim \{-\pi/6, -1\}$. Since we are forced to rationalize the fitted coefficients matrix inversion might never recover the correct result.

The second problem is that one can only invert a square matrix. So one must have exactly as many points $\bx_j$ as basis functions $\cB_i$ ($p=n$). As we will discuss, lattice reduction lets us trade digits of precision for additional points, but this is impossible with matrix inversion. One could attempt variations on matrix inversion, like a least-squares fit where you try to find the best (real number) $c_i$ with the information from $p$ points. However, if $p<n$ the result is undetermined and the least-squares fit may not give the correct answer. A least-squares fit also cannot exploit that $c_i$ are rational. 

 Another problem with matrix inversion (which as we will see persists in other methods) is that the precision on the $c_i$ after this procedure can be much lower than the precision of $M$ and $f(\bx_i)$. To leading order \cite{Cheney:2007}
 \begin{equation}
 \frac{\|\Delta c\|}{\|c\|}
\;\approx\;
\kappa(M)\,\Bigl(\frac{\|\Delta f\|}{\|f\|} + \frac{\|\Delta M\|}{\|M\|}\Bigr)\,, \label{eq:errorprop}
\end{equation}
where $\|\cdot\|$ is a norm and $\kappa$ is the condition number defined by
\begin{equation}
    \kappa(M) = \| M \| \| M^{-1} \| \,.
    \label{eq:condnum}
\end{equation}
Eq.~\eqref{eq:errorprop} says that the uncertainty on the outputs is given by the uncertainty of the inputs amplified by the condition number. 

If the entries are all the same order of magnitude and normally distributed, then the condition number scales like the dimension of $M$: $\kappa \sim n$ at large $n$~\cite{Edelman1988}. 
When the entries are not uniform, then the condition number is larger.
For example, if $M$ is diagonal with $a < M_{ii} < b$ then $\kappa = \frac{b}{a}$. Our case, where the entries come from the evaluation of smooth functions is actually much worse. In the limit that two points are coincident, the matrix is not invertible and the condition number is infinite. If the points are close but not coincident, the conditional number can be enormous. As we will see in Section~\ref{sec:lll_limitations}
this is in fact what happens in our case: the condition number is generally exponentially large $\kappa\sim 10^n$ instead of $\kappa\sim n$ for random matrices. 
That analytic regression of the coefficients of smooth functions is an ill-conditioned problem is a common challenge for  all the methods we will discuss, but it is particularly debilitating for matrix inversion. We return to this topic in Section~\ref{sec:lll_limitations}.

\subsection{Lattice reduction}

Lattice reduction offers the advantage over matrix inversion of being able to both exploit that the $c_i$ are rational numbers and to vary the number of points used for the regression.

The goal of a lattice reduction algorithm is to reduce an initial basis into an equivalent one with shorter vectors by some measure. For high-dimensional lattices, this task is challenging. For instance, exact algorithms are able to recover the shortest vector in the lattice, but come at a cost of exponential time complexity in the lattice dimension. Instead,  one has to rely on approximation algorithms, trading exactness of the result for a more reasonable runtime complexity. The classic polynomial-time algorithm for lattice reduction is called the LLL (Lenstra–Lenstra–Lovász) algorithm \cite{Lenstra:1982}. 
Even though LLL is not guaranteed to recover the optimal lattice, the length of the shortest vector after lattice reduction is at least an upper bound on the length of the absolute smallest possible lattice vector. This will be enough for our purposes. 
Various algorithms other than LLL are widely used. For example, \textsc{Mathematica} uses the L$^2$ algorithm~\cite{Nguyen:2001}. We give a brief overview of some lattice reduction algorithms in Appendix~\ref{app:lll_details}. 

We will use lattice reduction to find integer relations between the basis functions and the true function. Other algorithms for finding relations amongst integers exist and have already seen widespread usage in physics, amongst which the PSLQ algorithm \cite{Ferguson:1992} is a competitive option. PSLQ has already been extensively used in computing Feynman integrals, for instance to fit $g-2$ contributions \cite{Laporta:2017okg}, or to recover analytical expressions from high-precision numerical evaluations of Feynman integrals at boundary points \cite{Chachamis:2008fx,Lee:2010cga,Chicherin:2017dob, Lee:2018ojn, Agarwal:2021zft, Canko:2024ara}. Other usage examples include deriving relations among multiple zeta values \cite{Acres:2021sss}. We note however that PSLQ is not a lattice reduction algorithm and only works to find relations between a set of real numbers, not lattice vectors.  In our work we will be leveraging evaluations at many kinematic points, which cannot be done for PSLQ. However, if we take only a single point ($p=1$) we can still use PSLQ to fit entire set of functions if enough digits of precision are available. 

To use lattice reduction to solve Eq.~\eqref{eq:problem_statement}, we first pick a set of $p$ evaluation points $\left\{\mathbf{x}_1, \cdots, \mathbf{x}_p \right\}$ to evaluate $f(\bx)$ and the $\cB_j(\bx)$. We also need the exponent of the multiplier $10^s$ used to compute the initial integer lattice.With these inputs, we construct a matrix of the form
\begin{equation}\label{eq:mat_lll}
M= \text{round } 10^s  \left(\begin{array}{c|c}
        f(\mathbf{x_1}) \cdots f(\mathbf{x_p})& \multirow{4}{*}{\phantom{$\cdot$} $ 10^{-s} \, \mathbb{I}_{n+1}$} \\
        \mathcal{B}_1(\mathbf{x_1}) \cdots \mathcal{B}_1(\mathbf{x_p})&\\
        \vdots&\\
        \mathcal{B}_n(\mathbf{x_1}) \cdots \mathcal{B}_n(\mathbf{x_p})&
    \end{array}\right) \,.
\end{equation}
The left part of the matrix is obtained by multiplying the numerical evaluations by $10^s$ and rounding the resulting number so that the entries only involve integers. The right part of the matrix is obtained by then concatenating an $n+1$ dimensional identity matrix. The rows of $M$ can be viewed as lattice basis vectors $\mathbf{v}$ of dimension $n+p+1$ with the full lattice being defined by $n+1$ basis vectors.

The integer $s$ used should preserve the information of the function values. Suppose we know all the function values to a minimum number of significant digits $d$. It is also important to take into account the absolute size of the function values. Suppose the range of magnitudes of all the $f(\bx)$ and $\cB(\bx)$ values in the matrix are between $10^{\Delta_{\text{max}}}$ and $10^{\Delta_{\text{min}}}$.  Then in order not to throw out the information in the smaller numbers we should expect to need $d \ge {\Delta_{\text{max}}}-{{\Delta_{\text{min}}}}$. Then we want to choose $s$ to preserve all of the significant digits of the largest numbers when they are rounded. So we take 
\begin{equation}
    s=d- \Delta_{\text{max}}\,.
    \label{sformula}
\end{equation}
This choice ensures that all digits entering the reduction matrix $M$ will be accurate, while discarding the minimal amount of information.\footnote{For instance, for $d=4$ and values (0.1234, 6.789, 0.006938) we would retain $s=3$, giving (123, 6789, 7) as inputs.}

Lattice reduction then reduces this matrix to another $(n\!+\!1)\times(n\!+\!p\!+\!1)$ matrix which minimizes some measure of the size of the elements in the matrix. Pseudocode for LLL is given in Appendix~\ref{app:lll_details}. Briefly, the output relevant for our purposes is that the first row of the reduced matrix should give the lattice point with the smallest possible length. Although practical lattice reduction algorithms are not guaranteed to find the absolute minimum length vector, they often do and in practice one can adjust the points selected and precision to look for stability and confirmation.

Let $\mathbf{u}_1$ be the first row of the reduced matrix, purported to have the smallest possible norm. This new basis vector can by construction be described in terms of the old basis vectors so that we have $\mathbf{u}_1= \sum_{i=1}^{n+1} q_i \mathbf{v}_{i}$. Since $M$ was formed with an identity matrix, it is simple to find those coefficients. Decomposing $\mathbf{u}_1=(\mathbf{u}_L, \mathbf{u}_R)$ with $\mathbf{u}_L = (u_1, \cdots, u_p)$ and $\mathbf{u}_R = (u_{p+1}, \cdots, u_{n+p+1})$ we can check that all the coefficients are given by $q_i = u_{p+i}$ and 
\begin{equation} \label{eq:lin_rel_u1}
    \mathbf{u}_1 = \sum_{i=1}^{n+1} \, u_{p+i} \mathbf{v}_i \,.
\end{equation}
In deriving Eq.~(\ref{eq:lin_rel_u1}) we have only used information from the $\mathbf{u}_R$ part of the vector, so that if we now focus on the first $p$ entries we can read off the non trivial relations 
\begin{equation}
   10^{-s}  \begin{pmatrix}
        u_1 \\
        \vdots\\
        u_p
    \end{pmatrix}=  u_{p+1} \begin{pmatrix}
        f(\mathbf{x}_1) \\
        \vdots\\
        f(\mathbf{x}_p) 
    \end{pmatrix} +   \sum_{j=1}^{n} u_{p+j+1} \begin{pmatrix}
        \mathcal{B}_j(\mathbf{x}_1) \\
        \vdots\\
        \mathcal{B}_j(\mathbf{x}_p)
    \end{pmatrix} \,.
\end{equation}
Provided that the lattice reduction was successful, the integers $u_i$ should be small, so that, up to $\mathcal{O}(10^{-s})$ corrections, we have derived 
\begin{equation}
 \begin{pmatrix}
        f(\mathbf{x}_1) \\
        \vdots\\
        f(\mathbf{x}_p) 
    \end{pmatrix} = -   \sum_{j=1}^{n} \frac{u_{p+j+1}}{u_{p+1}} \begin{pmatrix}
        \mathcal{B}_j(\mathbf{x}_1) \\
        \vdots\\
        \mathcal{B}_j(\mathbf{x}_p)
    \end{pmatrix} + \mathcal{O}(10^{-s})\,,
\end{equation}
    which relate the unknown function in terms of the known basis functions. Since the $u_i$ are integers we can check that, as required, these relations only involves rational coefficients. If the fit is accurate, that is if the precision was sufficient and if the evaluation points were informative enough, we can then infer the general functional relation, valid for any point $\mathbf{x}$, as simply
\begin{equation}
    f(\mathbf{x})= - \sum_{j=1}^{n} \frac{u_{p+j+1}}{u_{p+1}} \mathcal{B}_j(\mathbf{x}) \,.
\end{equation}
An example is given in Eq.~\eqref{eq:llexample} and a more detailed illustration in Appendix~\ref{sec:app_simple_example_lll}.

\section{Examples}\label{sec:simple_integrals}
In this section we will demonstrate the analytic regression of Feynman amplitudes through a series of examples of increasing complexity.
  
We begin with the one-loop triangle with two equal masses, which corresponds to setting $z=1-\zb$ in Eq.~\eqref{eq:zform}, which has 17 basis functions.  In this example, there is a single kinematic variable and all the basis functions are classical polylogarithms easily evaluated to arbitrary accuracy.
We next look at some 2-loop integrals. We choose some that are functions of 2 variables to illustrate the extra complexity for a larger alphabet. The alphabet for these diagrams can be computed automatically, but constructing the basis functions and the algebraic prefactor requires some care and illustrates other technical elements of the procedure.
When multiple kinematic variables are relevant, then it is greatly beneficial to find set of integrable symbols before integrating them into GPLs. 
The trade-off between the number of sampled points and the evaluation precision is also discussed for this example and highlighted in Fig.~\ref{fig:1lfitresult}. 
Finally, we give two additional challenging examples: the triangle ladder diagrams up to three loops, in Section~\ref{sec:ladders} and in Section~\ref{sec:dbox}, the 2-loop outer-mass double box computed in~\cite{Caron-Huot:2014lda} and bootstrapped in~\cite{Hannesdottir:2024hke}.
With $n$ functions, the time to run lattice reduction generally scales like $t\sim n^4$ (see Section~\ref{sec:timing_scaling}) so it is desirable to keep the basis set as small as possible.
Linear combinations of basis functions can be eliminated with Landau bootstrap techniques (see~\cite{Hannesdottir:2024hke} for example). Note that in this section, we also make use of the C++ implementation of lattice reduction, \textsc{Fplll}~\cite{fplll}, to speed up the analytic regression.

\subsection{Example 1: The one-loop triangle}\label{sec:one_loop_example}
\unitlength1cm
The 1-loop massless triangle scalar Feynman integral is finite in $d=4$:
\begin{equation}\label{eq:1ltriangle}
    T=\triangleoneloop{p_1+p_2}{p_1}{p_2} \,.
\end{equation}
We take $p_1^2=p_2^2=-m^2$ for simplicity, so the amplitude is a function of $m$ and $s=(p_1+p_2)^2$.
Thus this loop is a special case of Eq.~\eqref{eq:zform}.
The integral is most naturally described using the variable \cite{Anastasiou:2006hc,Gehrmann:2013cxs}
\begin{equation}\label{eq:x_def}
    x=\frac{1-\sqrt{1+\frac{4m^2}{s}}}{1+\sqrt{1+\frac{4m^2}{s}}}\,.
\end{equation}
Running \textsc{SOFIA}~\cite{Correia:2025yao} we find 3 relevant letters: $\{x, 1+x , 1-x\}$, which in turn informs us that the candidate basis function set is given by 
\begin{equation} \label{eq:basis1ltriangle}
    \tilde{\mathcal{B}}^{(0)}  = \{G(a_1, a_2; x) , \pi G(a_3;x), G(a_4;x), \pi , \zeta_2 \}\,,
\end{equation}
where $a_i = 0, \pm 1$, for a total of  $n=17$ basis functions $\tilde{\cB}_j$. Here we retained all functions up to transcendental weight 2. In this case, the GPLs can all be written as classical polylogarithms (e.g. $G(1,1;x) =\frac{1}{2}\ln^2(1-x)$). 

\textsc{SOFIA} also gives us the maximal cut $P(x)$, allowing us to parametrize the amplitude's analytical expression as\footnote{To ensure that we have the correct factors of $\pi$ in $P(x)$, we select a particular point, $x=1/2$, and do a PSLQ fit. The fit relates the basis functions $\{\mathcal{B}_i(1/2)\}$ and different prefactors $\{\pi^\alpha P(1/2)\}$ where we vary $\alpha \in [-5,5]$. Our fit is successful and assigns a non-zero coefficient only for the $\alpha=0$ entry.}
\begin{equation}
    T(x) = P(x) \sum_{i=1}^{17} c_i \, \tilde{\mathcal{B}}_i (x) \quad \text{with} \quad P(x)=\frac{x}{m^2(1-x^2)}
\end{equation}
as in Eq.~\eqref{BBform}.

Using \textsc{AMFlow} with the IBP package \textsc{Blade}, we sample the amplitude $T(x)$ at kinematic points randomly picked in $x\in[0,1]$. Each point takes around 6 CPU-min on a laptop with an Apple M1 chip for 30 significant digits. For the points sampled we have $0.7<|T(x)|<2550$, with median values around $T(x)\sim \mathcal{O}(1)$. 
Evaluating the basis of Eq.~\eqref{eq:basis1ltriangle}) on the same kinematic points (which is essentially instant since the basis functions are classical polylogarithms), we can then construct the matrix in Eq.~(\ref{eq:mat_lll}) and apply lattice reduction.
The result is 
\begin{equation}
    T(x) = \frac{x}{m^2 (1-x^2)} \left(-\frac{\pi^2}{3} + 4 G(0,-1,x) - 2 G(0,0;x)\right) \,,
\end{equation}
which is the correct answer (see \cite{Gehrmann:2013cxs} for example). 

\begin{figure}[t]
    \centering
    \hspace{2cm}
    \includegraphics[width=0.8\linewidth]{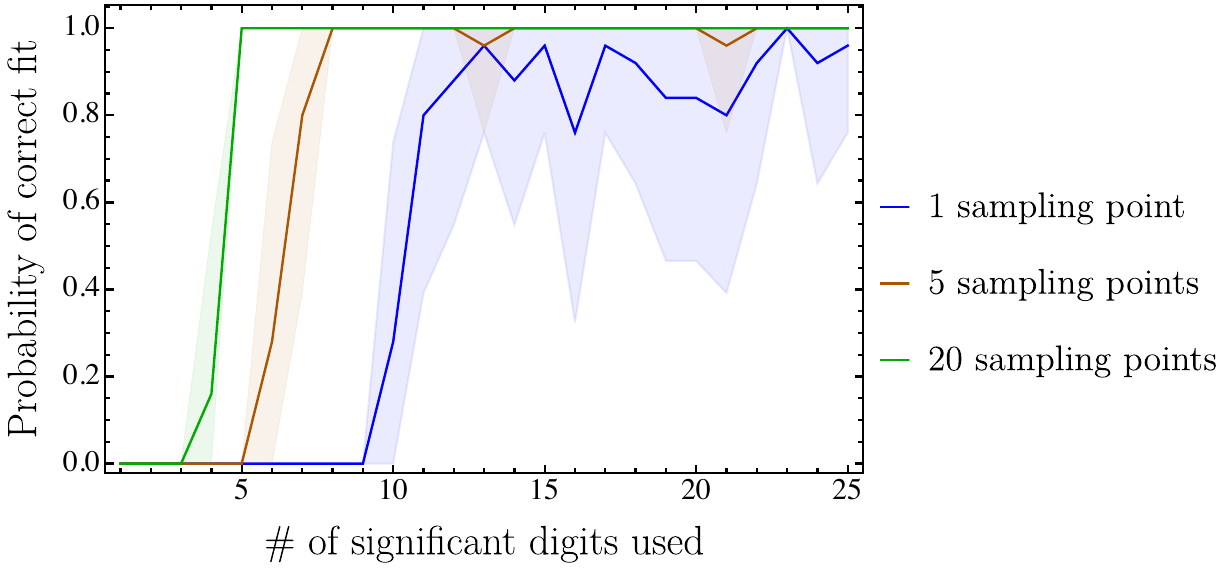}
    \caption{Success rate of our lattice fitting procedure for recovering the analytical expression corresponding to the 1-loop triangle of Eq.~(\ref{eq:1ltriangle}). The results are averaged over 25 trials, with the shaded band corresponding to the standard deviation.}
    \label{fig:1lfitresult}
\end{figure}

To calibrate the method, we vary the number of digits and points used in this example. The probability of recovering the correct analytical expression is displayed in Fig.~\ref{fig:1lfitresult}. 
We find that with a single point, one can often find the correct result with 12 or more digits, but not always -- there is a large variance based on which point we pick. With 5 points, having 8 digits of precision is almost always sufficient, and with 20 points, we can use 5 digits and get the right answer 100\% of the time.  
Thus, we see a clear advantage as we increase the number of sampling points. 

Compared to  matrix inversion, lattice reduction allows us to recover the correct relation even in the regime where we have fewer sampling points than basis functions. 
Even when using the same number of points as matrix inversion ($n=p=17$), lattice reduction only requires 5 digits of precision while matrix inversion requires at least 12 digits to guarantee success with a rationalization error of $10^{-2}$. 

Compared to PSLQ, which only works with a single point ($p=1$), lattice reduction is also far superior. With PSLQ we find that we need $21.4\pm 6.0$ digits of precision, with the spread representing the uncertainty based on the choice of point.  
If we use lattice reduction with a single point, we find that $10.8 \pm 0.6$ digits of precision are needed.

\subsection{Example 2: Master integrals for two-loop four-point diagram \label{sec:masters}}
Next, we consider a set of Feynman integrals that appear as master integrals in the reduction of two-loop four-point diagrams, as found in \cite{Gehrmann:2013cxs}:
\begin{equation}
\parbox{0.95\textwidth}{
 \triangletwosfourd{p_{12}}{q_2}{q_1}{I_1(x)} 
\triangletwosfived{q_2}{p_{12}}{q_1}{I_2(x)} 
\boxthreesfived{p_1}{p_2}{q_1}{q_2}{I_3(x,z)} 
\boxthreessevend{p_1}{p_2}{q_1}{q_2}{I_4(x,z)} 
 \label{fig:twoloop_fourpoint}
 \vspace{5mm}
}
\end{equation}
These master integrals are relevant for two-loop corrections to the calculation of vector boson pair production and are all expressible in terms of GPLs.

\subsubsection{Notation and methods}\label{sec:two_loops_methods}
 For the diagrams in Eq.~\eqref{fig:twoloop_fourpoint}, the incoming momenta involving adjacent legs are on-shell, satisfying $p_1^2=p_2^2=0$. 
 The other two legs are off-shell and the respective momenta satisfy $q_1^2=q_2^2=Q^2$ with $Q$ the mass of the vector boson. The dot in the first diagram $I_1$ indicates that the propagator is squared.
 These diagrams are all planar and the off-shell legs are adjacent. Rather than expressing results in terms of the usual Mandelstam variables $s=(p_1+p_2)^2$ and $u=(p_2-q_1)^2$, it is more convenient to use $(x,z)$, defined through
\begin{equation}\label{eq:two_loop_kinmatics}
    s=-m^2\frac{(1+x)^2}{x} \quad \text{and} \quad u=-m^2 z \,,
\end{equation}
where $m^2=-Q^2$. In particular, all master integrals will be real valued functions of $(x,z)$, provided we restrict them to the 
Euclidean region $z>0, 0<x<1, m^2>0$.\footnote{In terms of the original Mandelstam variables it corresponds to $s<0, u<0$ and $Q^2<0$. The result in the physical region can then be recovered by analytical continuation.} 
The diagrams in Eq.~\eqref{fig:twoloop_fourpoint} can all be described with the alphabet\footnote{Other topologies have additional Landau singularities and therefore require additional letters such as $1+x+x^2, 1+x+x^2+xz, x+z+x z+x^2 z$ and $x+(1\pm i \sqrt{3})/2$ to write the associated amplitudes. Even though these letters are not multilinear in $x$, the associated symbols can be directly integrated into their corresponding GPL forms provided the variable $z$ is integrated first. 
}
\begin{equation}\label{eq:alphabet_two_loop}
    \tilde{A} = \left\{x, z, 1+x, 1-x, 1-z, x+z, 1+x z\right\} \,.
\end{equation}
The integrals are calculated using dimensional regularization and expressed as a series in $\epsilon=(4-d)/2$ : 
\begin{equation}
    I(x,z) = P(x,z) \sum_{n=-n_1}^\infty \epsilon^n f_n(x,z) \,,
\end{equation}
where  $P(x,z)$ is the algebraic prefactor that is obtained through the maximal cut, $f_n(x,z)$ are GPLs and $n_1$ is the maximum degree of divergence in the dimensional regulator.  Since the numerical result of \textsc{AMFlow} is also given as a series in $\epsilon$ we will fit each of the functions $f_n$ independently\footnote{To make the connection with standard convention for Feynman integrals \cite{Weinzierl:2022eaz}, such as in Eq.~\eqref{eq:def_loop}, we will multiply the result given by \textsc{AMFlow} by the prefactor $e^{L \epsilon \gamma_E}$, with $L$ the number of loops. This ensures that the functions $f_n$ are free of $\gamma_E$ factors, resulting in a simpler result.}. Provided that $f_{-n_1}$ is given by GPLs of weights $n_\star$, we expect that each $f_n$ can be decomposed into a linear combination of GPLs of weights up to $n+n_1+n_\star$ \cite{Henn:2013pwa,Henn:2014qga}. In the following, we will be interested in recovering the analytical expressions corresponding to the amplitudes of Eq.~\eqref{fig:twoloop_fourpoint} up to $\epsilon^0$.

To construct the basis of GPLs we follow the procedure outlined in Section~\ref{sec:basis_set}, starting by writing down all possible integrable symbols with letters drawn from the alphabet of Eq.~(\ref{eq:alphabet_two_loop}). This procedure is best done iteratively (see Appendix~\ref{sec:symbols}), where integrable symbols at weight $n$ are obtained using the integrable symbols at weight $n-1$ \cite{Chavez:2012kn, Hannesdottir:2024hke}. The result is a set of symbols (here at weight 4 for concreteness)
\begin{equation} \label{eq:symbol_weight4}
    \mathcal{S} = \sum_{i=1}^{|\tilde{A}|}  c_{i_1,i_2,i_3,i_4} \, L_{i_1} \otimes L_{i_2} \otimes L_{i_3} \otimes L_{i_4} \,,
\end{equation}
where $L_i$ are the letters of the alphabet and the coefficients $c_{i_1,i_2,i_3,i_4}$ are rational. Since the alphabet is multilinear in $x,z$ these symbols can be directly integrated into GPLs. Our functional basis will  contain these GPLs, but must also be extended to account for transcendental pieces missed by the symbol. Indeed, the symbol is insensitive to multiple zeta values and powers of $\pi$ , in that $\mathcal{S}[\pi^m f(x,z)]=0$. Therefore, at transcendental weight $n$, we must also include all functions given by $\pi^{m_1} f_{m_2}(x,z)$ where $f_{m_2}$ is a GPL of weight $m_2$ and $m_1+m_2=n$. We note that this can additionally help fix branch cuts in our final fit. For instance, if we include $\log x$ in the basis set we would need to also have $\pi$ to fit a real function sampled in the domain $x<0$. In the following equation, we provide all possible combinations of transcendental functions and $\pi$, zeta values from weight-$0$ to weight-$6$, which are relevant for all the examples in this paper. 
\begin{align}\label{eq:basis_tower}
    \text{Weight-0:  }&\quad 1\notag\\
    \text{Weight-1:  }&\quad G(a_1,x),\,\, {\color{darkblue} \pi}\notag\\
    \text{Weight-2:  }&\quad G(a_1,a_2,x),\,\, {\color{darkblue} \pi\times G(a_1,x)},\,\, \zeta_2\notag\\
    \text{Weight-3:  }&\quad G(a_1,a_2,a_3,x),\,\, {\color{darkblue} \pi\times G(a_1,a_2,x)},\,\, \zeta_2\times G(a_1,x),\,\, \zeta_3\notag\\
    \text{Weight-4:  }&\quad G(a_1,a_2,a_3,a_4,x),\,\,{\color{darkblue} \pi\times G(a_1,a_2,a_3,x)},\,\, \zeta_2 \times G(a_1,a_2,x),\,\, {\color{darkblue} \pi^3\times G(a_1,x)},\,\,\notag\\
    &\quad \zeta_3\times G(a_1,x),\,\,  \zeta_4\notag\\
    \text{Weight-5:  }&\quad G(a_1,a_2,a_3,a_4,a_5,x),\,\, {\color{darkblue} \pi\times G(a_1,a_2,a_3,a_4,x)},\,\, \zeta_2\times G(a_1,a_2,a_3,x),\,\, \notag\\
    &\quad {\color{darkblue} \pi^3\times G(a_1,a_2,x)},\,\, \zeta_3\times G(a_1,a_2,x),\,\,  \zeta_4\times G(a_1,x),\,\, \zeta_5,\,\,\zeta_2\times \zeta_3\notag\\
    \text{Weight-6:  }&\quad G(a_1,a_2,a_3,a_4,a_5,a_6,x),\,\, {\color{darkblue} \pi\times G(a_1,a_2,a_3,a_4,a_5,x)},\,\, \zeta_2\times G(a_1,a_2,a_3,a_4,x),\,\,\notag\\
    &\quad {\color{darkblue} \pi^3\times G(a_1,a_2,a_3,x)},\,\, \zeta_3\times G(a_1,a_2,a_3,x),\,\, \zeta_4\times G(a_1,a_2,x),\,\, \zeta_5\times G(a_1,x),\,\,\notag\\
    &\quad \zeta_2\zeta_3\times G(a_1,x),\,\, {\color{darkblue} \pi^5\times G(a_1,x)},\,\,\zeta_6,\,\, \zeta_3^2\notag\\
    \cdots
\end{align}
where $G(a_1,\cdots,x)$ stands for all possible GPLs obtained from integrating the symbol basis.
Here, all the terms in black are naturally present in a standard Feynman integral, and all the {\color{darkblue} blue} terms, with odd powers of $\pi$, usually arise from the branch cuts of those GPLs. In other words, they are always accompanied by the complex number $i$, and are likely to be eliminated by rewriting the GPLs with appropriate identities. In practice, if all the GPLs we construct give real numbers in the domain in which we are working, it is safe to construct a function basis without them. In the following, we will refer to all possible terms (black and {\color{darkblue} blue}) as the full basis (`full'), the black terms as the simplified basis (`sim'), and the weight-$n$ line as the uniform basis (`unif').

\subsubsection{Results}
 For the first two diagrams of Eq.~\eqref{fig:twoloop_fourpoint} the only two scales in the problem are $p_{12}^2=s$ and $Q^2$ so that we expect the amplitude to only depend on $x$. Running \textsc{SOFIA} we find the alphabet for either of these two diagrams to be
\begin{equation}\label{eq:alphabet_two_loop_red}
    \tilde{A}_1 = \{ x, 1+x, 1-x\}\,,
\end{equation}
which is indeed a subset of Eq.~(\ref{eq:alphabet_two_loop}). 

Obtaining the maximal cut for the first diagram, labeled $I_1(x)$, requires some care~\footnote{The actual computation of the prefactor in \textsc{SOFIA} requires integrating over a remaining Baikov variable, where the integration contour is fixed by asking for the Gram determinant to be positive. Performing this for a generic integral can be involved but for the simple examples at hand we will assume that this procedure can be carried out, and we directly read-off the prefactors $P(x)$ from \cite{Gehrmann:2013cxs}.}. In this diagram the dot indicates an additional power of the propagator in the denominator. To account for it, we can follow Ref.~\cite{Primo:2016ebd} by turning the double propagator into a simple one and insert a fake mass $\tilde{m}$. Since $\lim_{\tilde{m}^2\rightarrow 0} \partial_{\tilde{m}^2} (\ell^2-\tilde{m}^2)^{-1} = \ell^{-4}$, it is sufficient to compute the cut for this alternate diagram, differentiate with respect to $\tilde{m}$ and send the fake mass to 0. This procedure determines the rational prefactor to be $P_1(x) = x  m^{-2} (1-x^2)^{-1}$. For the second diagram in Eq.~\eqref{fig:twoloop_fourpoint},  $I_2$ , the maximal cut can be directly evaluated and the prefactor is identical, with $P_2(x)=P_1(x)$. 

For the last two diagrams of Eq.~\eqref{fig:twoloop_fourpoint} we have three scales in the problem so we need the full dependency on $x$ and $z$. The associated maximal cuts can be directly computed and we retain $P_3(x,z)= m^{-2} (1-z)^{-1}$ and $P_4(x,z)=x^2 m^{-6} z^{-1} (1+x)^{-4}$. Running \textsc{SOFIA} for $I_3$ we indeed confirm that the alphabet of Eq.~(\ref{eq:alphabet_two_loop}) is the relevant one. When running \textsc{SOFIA} on $I_4$ we find two additional letters, $1+x+x^2+xz$  and $x+z+xz+x^2z$ on which the final answer does not depend. For simplicity, and to reduce computational costs, we discard these letters initially. If those letters were actually needed in the final analytical answer, we would expect our fits to fail, either predicting zero relations, or not being robust to a precision increase or a different point sampling.

We sample the various integrals in \textsc{AMFlow} with \textsc{Blade} as the IBP reducer. In the second column of Tab.~\ref{tab:two_loop_results} we catalog the time required for the numerical evaluation of a single kinematic point. We draw the kinematic variables with $x\in[0,1]$ and $z\in [0,1]$ since this domain guarantees that both the integral and the function basis will be real valued. While  lattice reduction can be used with complex numbers, we found that retaining only real numbers leads to a speed-up and is sufficient for the examples at hand. As described in Section~\ref{sec:two_loops_methods}, we construct the basis of GPLs from the integrable symbols and possible powers of $\pi$, keeping only the functions required at each order in the $\epsilon$ expansion. For instance, the integral $I_1(x)$ has contributions at $\epsilon^{-1}$ and $\epsilon^{0}$. In this example, the $\epsilon^0$ piece has transcendental weight 3, so that we only need a basis set up to transcendental weight 2 functions to fit the $\epsilon^{-1}$ contribution. The set has 17 functions and turns out to be equivalent to the $\tilde{\mathcal{B}}^{(0)}$ set of Eq.(\ref{eq:basis1ltriangle}), as expected, since the alphabets are identical. In contrast, the weight 3 basis set required for the $\epsilon^0$ part has 58 relevant functions and reads 

\begin{equation}\label{eq:basis2ltriangle}
  \tilde{\mathcal{B}}^{(1)}  = \{ G(a_1, a_2, a_3; x),  \pi G(a_4, a_5; x) , \pi^2 G(a_6;x), \zeta_3, \pi^3\} \cup \tilde{\mathcal{B}}^{(0)} \,,
\end{equation}
with $a_i=0, \pm 1$. The size of all the relevant basis sets are given in the fifth column of Tab.~\ref{tab:two_loop_results}. We emphasize that for $I_1$ and $I_2$ the relevant alphabet Eq.~(\ref{eq:alphabet_two_loop_red}) has only 3 letters, compared to the 7 of Eq.~(\ref{eq:alphabet_two_loop}), which helps to reduce the size of the basis set.

\begin{table}[t!]
    \centering
    \begin{small}
\begin{tabular}{l >{\centering\arraybackslash}p{2cm} >{\centering\arraybackslash}p{2cm} >{\centering\arraybackslash}p{2.5cm} >{\centering\arraybackslash}p{1.5cm} >{\centering\arraybackslash}p{2cm}}
    \toprule
\bfseries Diagram & \bfseries   \textsc{AMFlow} point time &\bfseries    \# points sampled  &\bfseries   Expansion order/weight &\bfseries   Basis size  &\bfseries   Reduction time    \\
\midrule
 \multirow{2}{*}{Triangle $I_1(x)$} &   \multirow{2}{*}{8.5 CPU-min}&   \multirow{2}{*}{58} & $\epsilon^{-1}$/ 2  (full)& 17& < 1s  \\
& & & $\epsilon^{0}$/ 3 (full) &58& 4s \\
\midrule
 \multirow{2}{*}{Triangle $I_2(x)$} &   \multirow{2}{*}{22.4 CPU-min}&   \multirow{2}{*}{362} & $\epsilon^{0}$/ 4 (full) & 182& 32s  \\
 & & & $\epsilon^{0}$/ 4 (unif) &124& 11s \\
 \midrule
 \multirow{3}{*}{Box $I_3(x,z)$} &   \multirow{3}{*}{21.2 CPU-min}&   \multirow{3}{*}{99} & $\epsilon^{-2}$/ 1 (full) & 8& <1s  \\
 & & & $\epsilon^{-1}$/ 2 (full) &53& < 1s \\
 & & & $\epsilon^{0}$/ 3 (full) &274&  40 min \\
  \midrule
   \multirow{6}{*}{Box $I_4(x,z)$} &   \multirow{6}{*}{5.4 CPU-h}&   \multirow{6}{*}{332} & $\epsilon^{-4}$/ 0 (full) & 1& <1s  \\
 & & & $\epsilon^{-3}$/ 1 (full) &8& < 1s \\
 & & & $\epsilon^{-2}$/ 2 (full) &53& < 1s\\
 & & & $\epsilon^{-1}$/ 3 (full) &274& 6 min\\
 & & & $\epsilon^{-1}$/ 3 (galois) &141& 16 s\\
 & & & $\epsilon^{0}$/ 4 (galois) &633& 9.4 h\\
  \bottomrule
\end{tabular}
\end{small}
    \caption{Timing studies for recovering the analytic expressions of the two-loop diagrams of Eq.~\eqref{fig:twoloop_fourpoint}. The lattice reduction fits are all successful and done using 28 significant digits of precision. At a given transcendental weight, we indicate whether our basis set includes all lower transcendental functions (full) or whether it is uniform (unif). For the box $I_4$ diagram, we also look at basis sets that are uniform with the associated symbols invariant under $x\rightarrow 1/x$ (galois). For $I_4$ the weight 4 fit is only successful when using the galois basis.}
   \label{tab:two_loop_results}
  \end{table}

For certain integrals we expect the result to have uniform transcendental weight (see~\cite{Arkani-Hamed:2010pyv} for example). In the case of $I_2$ we can see  from Tab.~\ref{tab:two_loop_results} that assuming uniform transcendental weight leads to a basis that is $30\%$ smaller, with the lattice reduction step taking a third of the time (consistent with the $t\sim n^4$ scaling, as $0.70^4=0.34$). For the box integral $I_4$, we find it necessary to impose additional constraints in order to fit the $\epsilon^0$ contribution. Our starting point is to observe (numerically) that the limit $x\rightarrow 1$ of the $\epsilon^0$ piece is well-defined. From the definition of $x$ in Eq.~(\ref{eq:x_def}) we can see that this corresponds to a square root singularity, which will not be $\alpha-$positive \cite{Hannesdottir:2024hke}. To cancel this square root branch cut, the integral (and in particular its symbol) needs to be \textit{Galois even}, that is invariant under a sign flip of the square root. In turn, this implies that the symbol of $I_4(x,z)$  should be invariant under the inversion $x\rightarrow 1/x$. Accounting for this constraint places a restriction on the GPLs of weight 4 that can enter the basis set. Along with a uniform transcendentality assumption, this allows us to half the basis size, reducing it from 1282 functions to 633. We stress that without this restriction the lattice reduction fit would need either more precision or additional smartly sampled points in order to be successful. Even then, we would expect the fit time to be an order of magnitude higher, on account of the polynomial scaling of the lattice reduction time.

With our basis in hand, we evaluate it on the sampled kinematic points and follow the lattice reduction algorithm outlined in Section~\ref{sec:lll_algo}, using 28 digits of precision. To verify the reliability of our fits we check whether the result is stable as the number of digits of precision is slowly increased. Additionally we also compare our analytical expressions to the ones of \cite{Gehrmann:2013cxs}, after performing the appropriate $\epsilon$ expansion and accounting for differences in conventional factors of $\pi$. Our expressions are all equivalent, although they are sometimes given as rewritings of the formulas of \cite{Gehrmann:2013cxs}. This is due to our choice of basis. Since GPLs satisfy various polylogarithmic identities, there are many equivalent rewritings for the same amplitude.

\begin{figure}
    \centering
    \hspace{2cm}
    \includegraphics[width=0.8\linewidth]{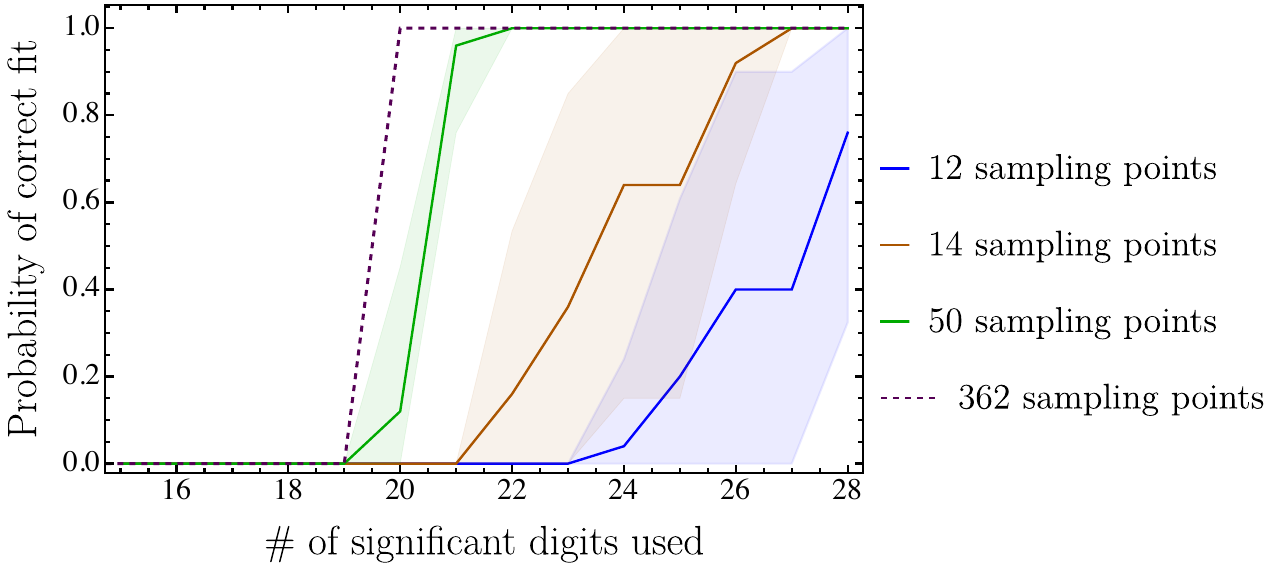}
    \caption{Success rate of our lattice reduction fitting procedure for recovering the analytical expression corresponding to the 2-loop triangle $I_2(x)$, depicted in Eq.~\eqref{fig:twoloop_fourpoint}, using a basis of 124 functions of uniform transcendental weight 4. The results are averaged over 25 trials, with the shaded band corresponding to the standard deviation. Each trial corresponds to using a different subset of the full 362 points available. The fit results using all available points are shown in dashed for reference.}
    \label{fig:2lfitresult}
\end{figure}

In summary, we successfully recovered all of the two-loop master integrals in Eq.~\eqref{fig:twoloop_fourpoint} using analytic regression with lattice reduction.  The step that requires the most analytical work is the construction of the basis. The numerical evaluations and lattice reduction, while time consuming, are completely automated and can be parallelized.
We found it was especially helpful to reduce the number of functions, especially in the regime where the number of precision digits is limited. As with the 1-loop example from 
Section~\ref{sec:one_loop_example}
we also explored the tradeoff between number of points and number of digits needed for a successful regression. Results are shown in  Fig.~\ref{fig:2lfitresult} for the integral $I_2(x)$, using the uniform basis of 124 functions. We see that using more points helps, but saturates: when we go from 50 points to 362 points there is little additional improvement.  This saturation effect can be understood analytically and is  discussed in Section~\ref{sec:lll_limitations}.

\subsection{Example 3: Triangle ladder diagrams to three-loops \label{sec:ladders}}

In this subsection, we consider the three-mass triangle ladder diagrams from one to three loops:
\begin{equation}
    \centering
     \triangleonel{p_3}{p_1}{p_2}
    \triangletwol{p_3}{p_1}{p_2}
    \trianglethreel{p_3}{p_1}{p_2}
    \vspace{0.5cm}\label{fig:threemass_triangle}
\end{equation}
We take generic external momenta $p_1^2\neq p_2^2 \neq p_3^2\neq 0$ with massless internal lines.

Using \textsc{SOFIA}, we can identify the following square root
\begin{equation*}
    \sqrt{(p_1^2)^2+(p_2^2)^2+(p_3^2)^2-2p_1^2p_2^2-2p_1^2p_3^2-2p_2^2p_3^2}\,,
\end{equation*}
and rationalize it with the conformal ratio variable $\{z,\bar z\}$:
\begin{equation}
    z\bar z = \frac{p_2^2}{p_1^2},\quad (1-z)(1-\bar z) = \frac{p_3^2}{p_1^2}\,.
\end{equation}
For simplicity, we will set $p_1^2=1$ and its explicit dependence can be recovered by dimensional analysis at the end. From the Landau analysis with \textsc{SOFIA}, we obtain the following alphabet for one- and two-loop triangles:
\begin{equation}
    A_{1,2}=\left\{z\bar z,\, (1-z)(1-\bar z),\, z-\bar z,\, \frac{\bar z}{z},\, \frac{1-z}{1-\bar z} \right\}\,.
\end{equation}
At three-loop, \text{SOFIA} gives 11 additional letters. To reduce the number of bases at weight-6 and computational costs, we will use a simplified alphabet \begin{equation}
    A_{3}^\star =\left\{z\bar z,\, (1-z)(1-\bar z),\, \frac{\bar z}{z},\, \frac{1-z}{1-\bar z} \right\}\,,
\end{equation} 
which turns out to be sufficient. Notice that we also remove the letter $z-\bar z$ in $A_3^\star$ since it does not appear at one-loop and two-loop. In practice, if one fails to bootstrap the integral with $A_3^\star$, one can gradually add back other letters and repeat the calculation until it works. 

From numerical evaluations, the triangle ladder diagrams are finite at $d=4$ and thus we do not need to perform any $\epsilon$ expansion. Similarly, we construct the integrable symbols up to transcendental weight $2L$ and use \textsc{PolyLogTools} to analytically integrate the symbol into GPLs and obtain the full function basis as in Eq.~\eqref{eq:basis_tower}. At one-loop, the letters in $A_{1,2}$ lead to 24 symbols, including both weight-1 and weight-2 after applying the integrability conditions. Integrating them to GPLs, we follow the construction in Eq.~\eqref{eq:basis_tower} and obtain 32 functions for the full basis, 26 for the simplified basis and 25 for the uniform weight-2 basis. Similarly, at two-loop, we have 300 integrable symbols, leading to 488 functions in the full basis, 393 in the simplified basis and 366 in the uniform weight-4 basis. As discussed above, we use a simpler alphabet $A_3^\star$ for three-loop computation, and the numbers are: 768 for the integrable symbols up to weight-6, 1373 functions for the full basis, 972 for the simplified basis and 806 for the uniform basis. It turns out that this alphabet is enough for bootstrapping the three-loop result, but in general, one can always extend it towards the full alphabet obtained from \text{SOFIA}.

Again we numerically evaluate the GPLs in the function basis using \textsc{GiNac} and the integrals using \textsc{AMFlow}. In this example, we use \textsc{Kira} together with \textsc{FireFly}~\cite{Klappert:2019emp,Klappert:2020nbg} to perform IBP reductions. We also randomly pick points in the unphysical region, $z,\,\bar z \in \mathcal{R}$, $0<z<\bar{z}<1$ for triangle ladders, such that all GPL bases are real.
After that, we construct the lattice matrix and perform the reduction using the C++ program \textsc{Fplll}. 
For one-loop and two-loop, we perform the reduction for all choices of basis: full, simplified and uniform. Note that in the full and simplified function basis, we do not assume any pre-knowledge of uniform transcendentality, but the lattice reduction still leads to that. For the three-loop, since the lattice reduction becomes very heavy, we only perform the analytic regression for the uniform case.
This gives rise to the following results:
\begin{align}
    T_1(z)&=\frac{1}{z-\bar z}\left[2\Li_2(z)-\Li_2(\bar z)+\log(z\bar z)\log\left(\frac{1-z}{1-\bar z}\right)\right]\,,\notag\\*
    T_2(z)&=\frac{1}{(1-z)(1-\bar z)(z-\bar z)}\Big[6\Li_4(z)-6\Li_4(\bar z)-3\log(z\bar z)\left(\Li_3(z)-\Li_3(\bar z)\right)\notag\\*
    &+\frac{1}{2}\log^2(z\bar z)\left(\Li_2(z)-\Li_2(\bar z)\right)\Big],\,\notag\\*
    T_3(z)&=\frac{1}{(1-z)^2(1-\bar z)^2(z-\bar z)}\Big[20\Li_6(z)-20\Li_6(\bar z)-10\log(z\bar z)\left(\Li_5(z)-\Li_5(\bar z)\right)\notag\\*
    &+\log^2(z\bar z)\left(\Li_4(z)-\Li_4(\bar z)\right)-\frac{1}{6}\log^3(z\bar z)\left(\Li_3(z)-\Li_3(\bar z)\right)\Big]\,.
\end{align}
In Tab.~\ref{tab:triangle_ladder_time}, we  show the time cost for \textsc{AMFlow} and \textsc{Fplll} for all the diagrams.

\begin{table}[t!]
    \centering
    \begin{small}
\begin{tabular}{l >{\centering\arraybackslash}p{2cm} >{\centering\arraybackslash}p{2.5cm} >{\centering\arraybackslash}p{2.0cm} >{\centering\arraybackslash}p{2cm} >{\centering\arraybackslash}p{2cm}}
    \toprule
\bfseries Diagram & \bfseries   \textsc{AMFlow} point time &\bfseries   Transcendental weights &\bfseries    \# points sampled   &\bfseries   Basis size  &\bfseries   Reduction time   \\
\midrule
 \multirow{3}{*}{One-loop} &   \multirow{3}{*}{15.6 CPU-min}&    \multirow{3}{*}{$\leq 2$} &  \multirow{3}{*}{5}    & full(32)& <1s  \\
& & & & simplified(26) & <1s \\
& & & & uniform(25) & <1s \\
\midrule
 \multirow{3}{*}{Two-loop} &   \multirow{3}{*}{1.1 CPU-h}& \multirow{3}{*}{$\leq 4$} & 100 &   full(488) & 9.6 min  \\
 & & & 100   & simplified(393) & 10.7 min \\
 & & & 60  & uniform(366) & 3.5 min \\
 \midrule
 \multirow{3}{*}{Three-loop} &   \multirow{3}{*}{5.7 CPU-h}& \multirow{3}{*}{$\leq 6$} &   - & full(1373) &  - \\
 & &  & - & simplified(972) &  - \\
 & &  & 200 & uniform(806) &  1.1 h \\
  \bottomrule
\end{tabular}
\end{small}
    \caption{Results for fitting the triangle ladder diagrams up to three-loops. For one-loop, the L${}^2$ algorithm only requires 3 significant digits, and for others, we use 20 significant digits. In the Basis size, the numbers inside the bracket represent the number of function bases with different settings. The calculation time is evaluated on Intel Core i7-12700K.
    For the three-loop triangle with the simplified or full basis, we did not find a solution with 400 points sampled and the computation cost becomes expensive.
    }
   \label{tab:triangle_ladder_time}
  \end{table}

\subsection{Example 4: The two-loop outer-mass double box} \label{sec:dbox}
For a final example, we consider the outer-mass double box integral
\vspace{5mm} 
\begin{equation}
\begin{gathered}
\outermassdbox{p_1}{p_2}{p_3}{p_4}{I_5(u,v)}
\end{gathered}
\label{fig:twoloop_dbox}
\end{equation}
\vspace{5mm} 

\noindent This diagram has already been computed, both through a direct computation relying on the method of differential equations \cite{Caron-Huot:2014lda} and via a bootstrapping of its symbol \cite{Hannesdottir:2024hke}. Although both methods produce the same result, they both require a significant amount of computational work even for experts.

\begin{figure}
    \centering
    \includegraphics[width=0.9\linewidth]{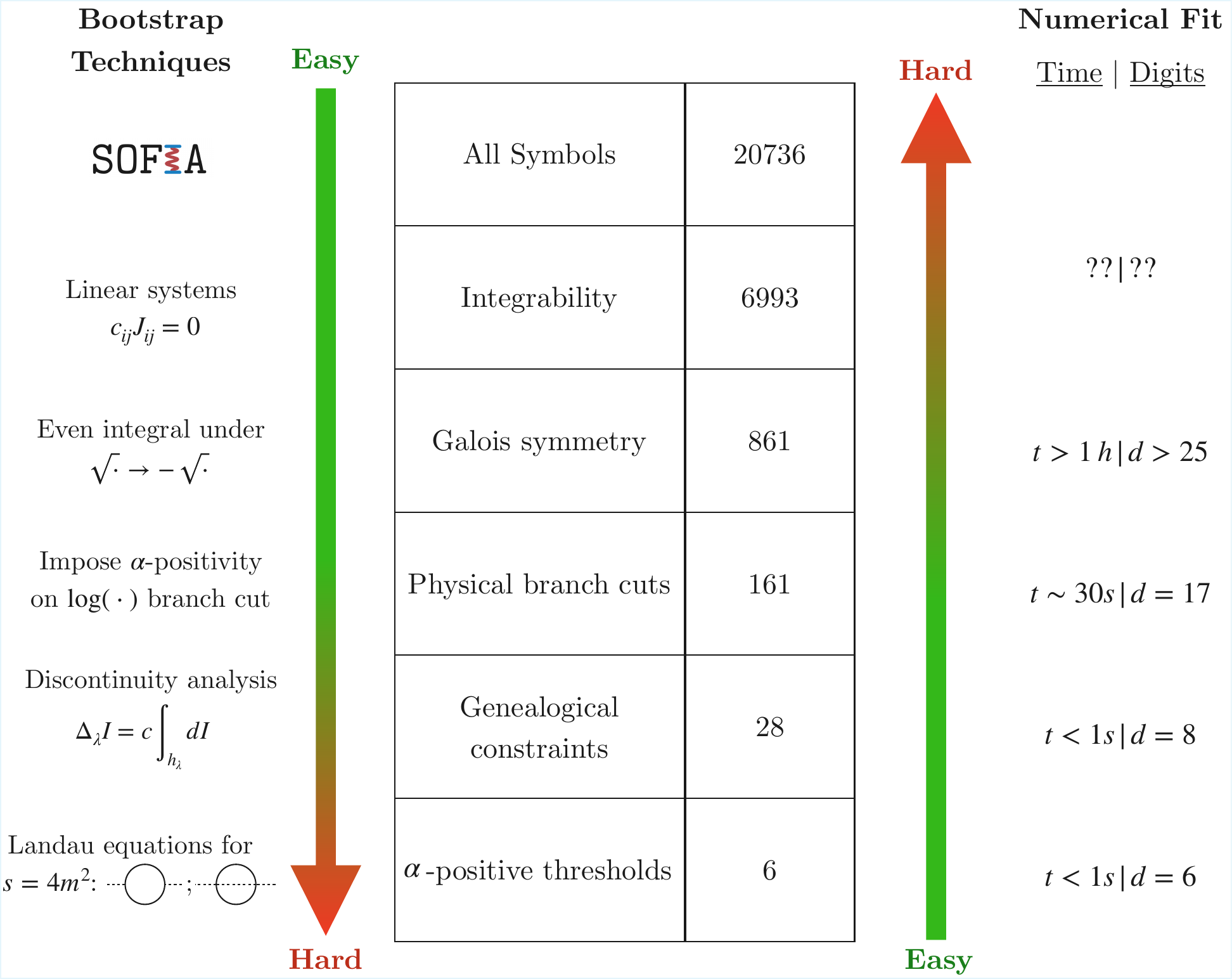}
    \caption{The outer-mass double box diagram can be inferred from two different directions. The top-down Landau bootstrap approach is analytical, while the bottom-up numerical approach fits the result using lattice reduction. Both approaches are complimentary in that they are easy to implement in opposing regimes.
}
    \label{fig:illustration_bootstrap}
\end{figure}

\subsubsection{Notation and methods}

The outer-mass double-box is a finite integral in four dimensions. Calling $m$ the common mass of the massive loop propagators, there are two dimensionless variables that the integral can depend on
\begin{equation} \label{eq:u_and_v_def}
    u = - \frac{4 m^2}{s} \quad \text{and} \quad v = - \frac{4 m^2}{t}\,. 
\end{equation}
Constructing the alphabet of the double box integral from the study of the Landau singularities leads to a set of 12 independent letters \cite{Hannesdottir:2024hke} given by 
\begin{align} \label{eq:alphabet_uv} \nonumber
    \tilde{A} = \Big\{& u, v, 1+u, 1+v, u+v, 1+u+v, \frac{\beta_u-1}{\beta_u+1} ,\\ & \frac{\beta_v-1}{\beta_v+1},\frac{\beta_{uv}-1}{\beta_{uv}+1},\frac{\beta_{uv}-\beta_u}{\beta_{uv}+\beta_u},\frac{\beta_{uv}-\beta_v}{\beta_{uv}+\beta_v} ,\frac{\beta_{uv}-\beta_u\beta_v}{\beta_{uv}+\beta_u\beta_v} \Big\} \, 
\end{align}
with $\beta_u = \sqrt{1+u}$, $\beta_v = \sqrt{1+v}$ and $\beta_{uv} = \sqrt{1+u+v}$. This set is over-complete for the purpose of writing down the final analytical expression, whose symbol depends only on 9 of those letters, but, a priori, the extra letters cannot be discarded. We assume the integral $I_5(u,v)$ is expressible as a universal prefactor $P_5(u,v)$ multiplying a linear combination of GPLs of uniform transcendental weight 4. (As before, if this assumption is too strong, we can consider multiple prefactors or lower weight GPLs.)
From the alphabet of Eq.~(\ref{eq:alphabet_uv}) there are  $12^4=20736$ possible symbol terms that one could write down, of which we can form linear combinations of 6993 integrable symbols. The bootstrap of the symbol of \cite{Hannesdottir:2024hke} proceeds by imposing additional constraints at the symbol level, such that only a single free parameter remains. In light of the results of Section~\ref{sec:simple_integrals}, it would be computationally expensive to perform a lattice reduction using the full basis of 6993 functions. We will instead employ a hybrid approach, combining some knowledge from the bootstrap with analytic regression and explore how many bootstrap constraints are necessary for the fitting procedure to be practical. The tradeoff is shown in Fig.~\ref{fig:illustration_bootstrap}.

\subsubsection{Function evaluation}
The numerical evaluation of $I_5$ is done in \textsc{AMFlow} with \textsc{BLADE} as the IBP reducer using a set of 200 kinematic points. We choose to sample our points in $u>0$ and $v>0$, restricting ourselves to the domain where the integral and the basis functions will be real-valued. Each integral point evaluation takes around 80 CPU-min on Intel Xeon Sapphire Rapids. We set the precision of \textsc{AMFlow} to 30 digits. 

\subsubsection{Constructing a set of basis functions}
To construct the basis functions we will take as starting point the sets of allowed symbols that remains after imposing various constraints. Following~\cite{Hannesdottir:2024hke} exactly, the successive constraints 
are integrability, galois symmetry, physical logarithmic branch cuts, genealogical constraints and asking for only algebraic $\alpha$-positive thresholds. The number of allowed symbols after imposing these constraints is respectively,  6993, 861, 161, 28 and 6. These symbols are originally expressed using the $u$, $v$ variables, which proves problematic when trying to integrate them analytically. Indeed, the letters of the alphabet of Eq.~(\ref{eq:alphabet_uv}) depend on various square roots, preventing a direct integration. To rationalize all of the square roots in the alphabet we perform a change of variables~\cite{Caron-Huot:2014lda}
\begin{equation}
    u = \frac{(1-w^2)(1-z^2)}{(w-z)^2} \quad \text{and}\quad v=\frac{4 w z}{(w-z)^2} \,,
\end{equation}
yielding a new set of letters
\begin{equation}
    \tilde{A}_2 = \left\{w, z, 1\pm w, 1\pm z, w \pm z, 1 \pm w z, 1 \pm w \mp z + wz\right\} \,,
\end{equation}
which are multilinear in the $w, z$ variables. In these new variables the symbols can be directly integrated and the associated GPLs can serve as a basis function set. Unfortunately, this approach suffers from two technical drawbacks. First, the \textit{FiberSymbol} routine can take hours to integrate a single symbol, especially if the expression has thousands of terms, which generically turns out to be the case here. Second, the integrated symbols are given by complicated GPLs expressions. While each entry in the basis of Eq.~(\ref{eq:basis1ltriangle}) and Eq.~(\ref{eq:basis2ltriangle}) was given by a single GPL, we now have entries that are composed of hundreds to thousands of terms. While \textsc{GiNaC} can still handle their numerical evaluations, the computational time grows significantly.

Instead of relying on an explicit analytical formula for the basis functions, it is also perfectly reasonable to define them using an integral representation that can only be done numerically. For instance, we can associate a numerical evaluation corresponding to the symbol of Eq.~(\ref{eq:symbol_weight4}) by the iterated integral over the various entries of the symbol
\begin{equation} \label{eq:iterated_int_symbol}
    \tilde{B}= \sum_{i=1}^{|\tilde{A}|}  c_{i_1,i_2,i_3,i_4} \int_0^1 d \log L_{i_4}(\lambda_4)  \int_0^{\lambda_4} d \log L_{i_3}(\lambda_3) \int_0^{\lambda_3} d \log L_{i_2}(\lambda_2) \int_0^{\lambda_2} d \log L_{i_1}(\lambda_1) \,,
\end{equation}
where the integration variables $\lambda_i$ are ordered and parametrize a path in kinematic space. In particular, the integral of Eq.~(\ref{eq:iterated_int_symbol}) is understood as having the kinematic variables appearing in the integrand defined as $(u(\lambda_j), v(\lambda_j))$. One simple choice of path is to take $(u(\lambda), v(\lambda))=(u/\lambda, v/\lambda)$ where $\lambda \in [0,1]$. This path satisfies the appropriate boundary conditions since as $\lambda \rightarrow 0$ we have $u, v \rightarrow \infty$, where the integral vanishes\footnote{From Eq.~(\ref{eq:u_and_v_def}) one can see that sending $u, v$ to infinity corresponds to giving the internal propagators infinite mass, such that the total integral vanishes.}. This choice of path is not unique. In particular, provided that the symbol is integrable, Eq.(\ref{eq:iterated_int_symbol}) will be independent of the specific path $(u(\lambda), v(\lambda))$ considered\footnote{We note, however, that for the purpose of constructing on over-complete basis imposing integrability is technically not essential. Indeed one could imagine fitting for the initial 20736 coefficients and expecting only a sparse subset of those to be non-zero, with the final answer depending only on linear combinations that correspond to integrable symbols.}. To implement Eq.~(\ref{eq:iterated_int_symbol}) numerically we have to address a couple of technical points. First, integrals where the letter $L_{i_1}(\lambda_1)$ has a simple pole at $\lambda_1=0$ are ill-defined. Typically one deals with this issue by introducing a tangential base point for defining the path (see section 5.2.5 of \cite{Brown:2013qva}) or by using a trailing zero prescription \cite{Walden:2020odh}, where one makes use of the shuffle product of iterated integrals. Analytically these procedures are well understood but in a pure numerical setting these techniques prove challenging to implement. The symbol bootstrap method precisely resolves this problem, since all symbols remaining after imposing physical logarithmic branch cuts will be free of such poles. Second, the high dimensionality of Eq.~(\ref{eq:iterated_int_symbol}) implies a high computational cost for obtaining increasing digits of precision. To facilitate the numerical evaluation we will then do the integral over $\lambda_1$ and $\lambda_4$ analytically, which trivially evaluate to logarithmic forms, reducing the problem to a two-dimensional numerical integral. We emphasize here that generically one would need to be wary of the boundary conditions considered, a point which we explore further in Appendix~\ref{sec:app_iterated_integrals}. It turns out that, for our problem, the expression of Eq.~(\ref{eq:iterated_int_symbol}) will be sufficient to fully represent the final answer.

\subsubsection{Results}
To recover the analytical formula for the double-box diagram of Eq.~\eqref{fig:twoloop_dbox} we will use the basis functions either explicitly given as GPLs or defined through their integral representation, contrasting and comparing both approaches.

We will be considering basis functions constructed from symbols where we have imposed (at least) integrability, galois symmetry and the presence of physical logarithmic branch cuts. All of these constraints are reasonably simple to implement and only require knowledge about the alphabet, the rational prefactor and the nature of the singularities. Running \textsc{SOFIA} and integrating over the remaining Baikov variable we find the prefactor to be 
\begin{equation}
    P_5(u,v)= \frac{u^2 v}{m^6 \sqrt{(u+1) (u+v+1)}} \leftrightarrow P_5(w,z)= \frac{4 w \left(w^2-1\right)^2 z \left(z^2-1\right)^2}{m^6 (w-z)^4 \left(w^2 z^2-1\right)} \,,
\end{equation}
which we need in both sets of variables. Indeed the GPL basis is naturally obtained in $(w,z)$ variables, while the 2-dimensional numerical iterated integral are easier in $(u,v)$ variables. Our successful fits are summarized in Tab~\ref{tab:two_loop_dbox_results}. For the GPL basis we can use \textsc{GiNaC} to get a numerical evaluation up to an arbitrary number of digits, retaining 30 digits to match the precision set in \textsc{AMFlow}. We will use 28 digits for the fits and verify the stability of the result with the extra digits of precision. Each evaluation of a basis function takes around 210 seconds on the set of 200 kinematic points. Naturally, one also has to account for the time taken to integrate the symbol, which can take many CPU hours per symbol. This is to be contrasted with the basis given by a numerical 2D integral. Using as working precision the machine precision, our integrals, evaluated in \textsc{Mathematica}, can be expected to be accurate only up to 16 digits, but require only 18 seconds to be evaluated on the same 200 points. Asking for more precision requires increasing the working precision which is computationally costly. Nevertheless, all of our fits are still accurate using only 15 digits, in part due to the extreme sparsity of the final answer\footnote{One may wonder whether using that same level of precision would be possible for the basis given by GPLs. As it turns out, using GPLs we require at least 17 digits to accurately fit the set of 161 basis functions. The discrepancy comes from the fact that the numerical integrals and the GPLs are not exactly the same basis functions. They differ by boundary terms that cancel in the final answer. The individual GPL basis on average evaluates to numbers that are $10^4$ larger, which hurts the fit and requires more precision to be resolved.}. We note that reaching a higher precision on the numerical integrals is simpler to achieve if an additional integral is performed analytically. The resulting one-fold integral representations are well suited for numerical evaluation, with the number of digits of accuracy increasing linearly with computing time \cite{Henn:2013pwa, Henn:2014qga}.

\begin{table}[t!]
    \centering
    \begin{small}
\begin{tabular}{>{\centering\arraybackslash}p{6cm} >{\centering\arraybackslash}p{1.5cm} >{\centering\arraybackslash}p{3.1cm} >{\centering\arraybackslash}p{2cm}}
    \toprule
\bfseries Extra Constraints imposed & \bfseries Basis size & \bfseries Basis type &\bfseries   Reduction time  \\
\midrule
 \multirow{2}{5.5cm}{\centering Physical branch cuts + genealogical + $\alpha$-positive thresholds} &   \multirow{2}{*}{6}& GPLs (28 digits)& < 1 s  \\
& &Integrals (15 digits) & < 1 s \\
\midrule
 \multirow{2}{5.5cm}{\centering Physical branch cuts + genealogical} &   \multirow{2}{*}{28}& GPLs (28 digits)& < 1 s  \\
& &Integrals (15 digits) & < 1 s \\
\midrule
 \multirow{2}{5.5cm}{\centering Physical branch cuts} &   \multirow{2}{*}{161}& GPLs (28 digits)&  40 s  \\
& &Integrals (15 digits) & 9 s \\
  \bottomrule
\end{tabular}
\end{small}
    \caption{Setting and timings for recovering the analytic expressions of the two-loop diagram of Eq.~\eqref{fig:twoloop_dbox}. The lattice reduction fits are all successful and use 200 different kinematic points. The fits are done using either 28 significant digits for the basis given as explicit GPLs or 15 digits for the basis given as numerical 2D integrals. The extra constraints are imposed at the symbol level, on top of integrability, and galois symmetry.}
   \label{tab:two_loop_dbox_results}
  \end{table}

Thus, it appears beneficial to use the numerical basis as opposed to the analytical one, since it is better behaved and simpler to evaluate. Our fitting procedure, coupled with the bootstrap of the symbol, allows for a complete fit using only minimal knowledge of the target integral. Each successive constraint facilitates the lattice reduction procedure, but one can already perform a valid fit using only symbols that correspond to physical branch cuts.  We stress, however, that we were able to use this reduced numerical basis directly due to the vanishing of all boundary terms. In a different setting, one may need to account for $\pi^2$ terms as we have done previously in Section~\ref{sec:simple_integrals}.

\section{Scaling behavior of lattice reduction}\label{sec:lll_limitations}

In Section~\ref{sec:simple_integrals} we presented a series of Feynman integrals which we were able to compute entirely by analytic regression. We found that generally the more points $p$ we used for the regression, the fewer digits of precision $d$ were needed. However, we also found that after reaching a critical number of evaluation points we would hit a \textit{precision plateau}. In this section we aim to make both these observations more precise, and discuss more generally the scaling behavior and possible limitations of  the lattice reduction approach.

\subsection{Precision scaling with number of points and basis functions}\label{sec:p_d_scaling}

To begin, let us try to understand the tradeoff between the number of points $p$ used and the number of digits of precision $d$ needed to fit $n$ basis functions in lattice reduction.

Suppose the coefficients in the true solution are rational numbers with numerator and denominator of order $10^R$. Lattice reduction will then fail if some linear combination of the basis functions with coefficients also of size $10^R$ evaluated at the points happens to vanish accidentally. To quantify this, consider first a single point ($p=1$). Lattice reduction first rescales and rounds the $n$ basis functions evaluated at the point into integers of size $10^d$ and combines then into a vector $\vec{u}$. If $\vec{c}$ is a vector of coefficients of size $10^R$ then a spurious solution will have $\vec{c} \cdot \vec{u}=0$. The number of $n$-vectors $\vec{c}$ with $|\vec{c}| < 10^R$ is around $10^{Rn}$, so there are around $10^{Rn}$ values for $\vec{c} \cdot \vec{v}$. Since each of these is a $d$ digit number, for 0 to be in the span of $\vec{c} \cdot \vec{v}$  requires on average that $10^{Rn} < 10^d$ so that $d> R n$. With more than one point we then require 
$\vec{c} \cdot \vec{v_j}$ to vanish for each point. But this is the same as the vanishing of a single point in $d\, \times\,n$ dimensions. So we conclude that we expect to need $d \ge \dmin$ digits of precision where
\begin{equation}
    \dmin \approx R \frac{n }{p} \,.\label{theory}
\end{equation}
Put simply,
the ``total digits'' of information available to the algorithm is $d\times p$ so an intelligent algorithm should be able to trade $d$ and $p$ holding their product fixed with comparable results.

\begin{figure}
    \centering
    \includegraphics[width=0.7\linewidth]{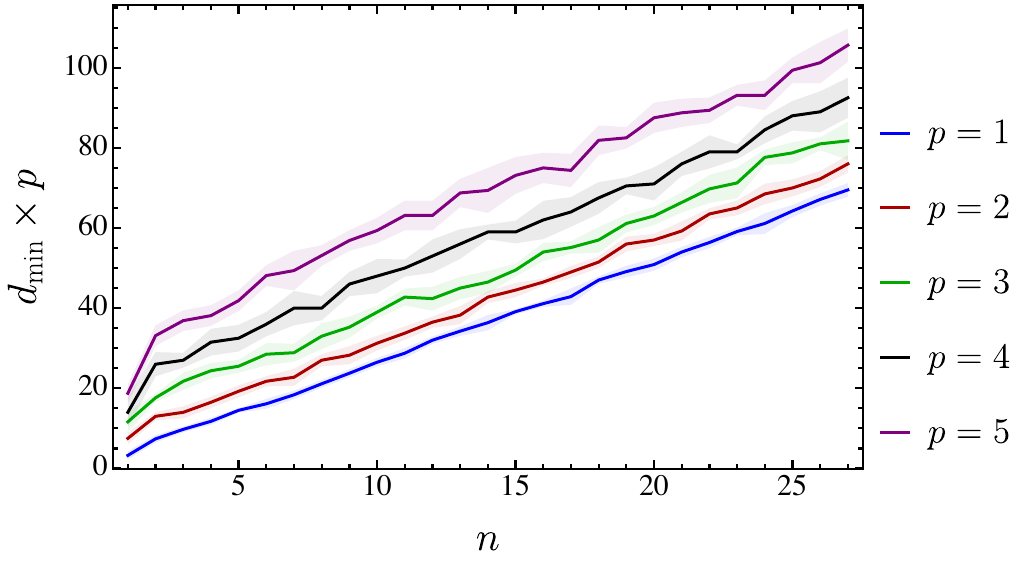}
    \caption{Minimum number of digits $\dmin$ required for successful lattice reduction multiplied by the number of points  $p$, shown as a function of the number of functions $n$ used. Weight 3 single variable GPLs are used for this figure evaluated at $p$ points $x$ with $-100 < x < 100$. The shaded bands reflect the uncertainty over 8 trials with different points. We see that  $\dmin \times p \sim n \times c$ for some constant $c$ which is similar for all values of $p$.
    }
    \label{fig:pxd}
\end{figure}

The behavior  $\dmin \times p \approx n$  is confirmed numerically in  Fig.~\ref{fig:pxd}. For this plot, the functions used are a subset of $n$ of the 27 weight-three single variable GPLs relevant for the triangle diagram of Section~\ref{sec:one_loop_example} 
multiplied by random integer coefficients between -100 and 100 (so $R=2$). 
This plot also shows that as the number of points grows there is an offset as well, so the fit is closer to 
\begin{equation}
    \dmin \approx \Reff \frac{n}{p}\ + d_0
    \label{numeric}
\end{equation}
for some $\Reff$ and $d_0$. In this plot $\Reff \approx 2.6$ which is close to $R=2$ and $d_0 \approx 5$. This numerical exercise and the fit in Eq.~\eqref{numeric} indicate that the theoretical result in Eq.~\eqref{theory} is quite accurate, but breaks down at large $p$; at large $p$, the number of digits needed saturates at $d_0$, leading to a plateau in $d$ even as more points are added.

\begin{figure}
    \centering
    \includegraphics[width=0.7\linewidth]{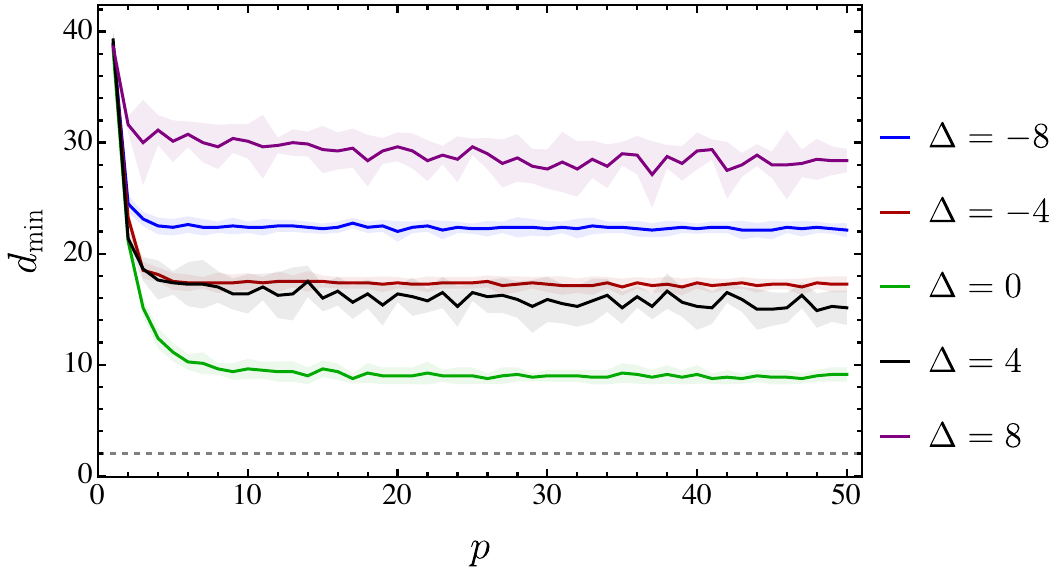}
    \caption{Minimum number of digits $\dmin$ required for successful lattice reduction shown as a function of the number of points $p$ used for different sampling ranges $\frac{1}{2}- 10^{\Delta} < x < \frac{1}{2}+ 10^{\Delta}$. The same functions are used as in Fig.~\ref{fig:pxd}. The shaded bands reflect the uncertainty over 8 trials with different points. The dashed gray line shows the minimum value of $d$, as dictated by the size of the coefficients in the true function. We see a plateau for $p > 6$ for all curves. For $\Delta < 0$ the value of the plateau monotonically decreases with $\Delta$, as it is partially caused by insufficient information in the sampling region. 
   }
    \label{fig:plateau}
\end{figure}

The plateau in the minimum number of digits needed is unpacked further in Fig.~\ref{fig:plateau}. 
For Fig.~\ref{fig:plateau} we fixed the number of functions to $n=15$ and varied the number of points and the domain where we select them. To systematically explore larger domains, we introduce a parameter $\Delta$ and sample points within $\frac{1}{2}-10^\Delta \le x \le \frac{1}{2}+10^\Delta$. Fig.~\ref{fig:plateau} shows that the minimum number of digits needed increases with large positive or negative $\Delta$, with the optimum having $\Delta \approx 0$. 
The behavior at large negative $\Delta$ is easy to understand -- if the points are too close together, more significant digits are needed to distinguish the function values. At large positive $\Delta$ the behavior is also sensible: the function values start to deviate by orders of magnitude so more digits are required to resolve their differences. 

\begin{figure}[t]
    \centering
    \includegraphics[width=0.9\linewidth]{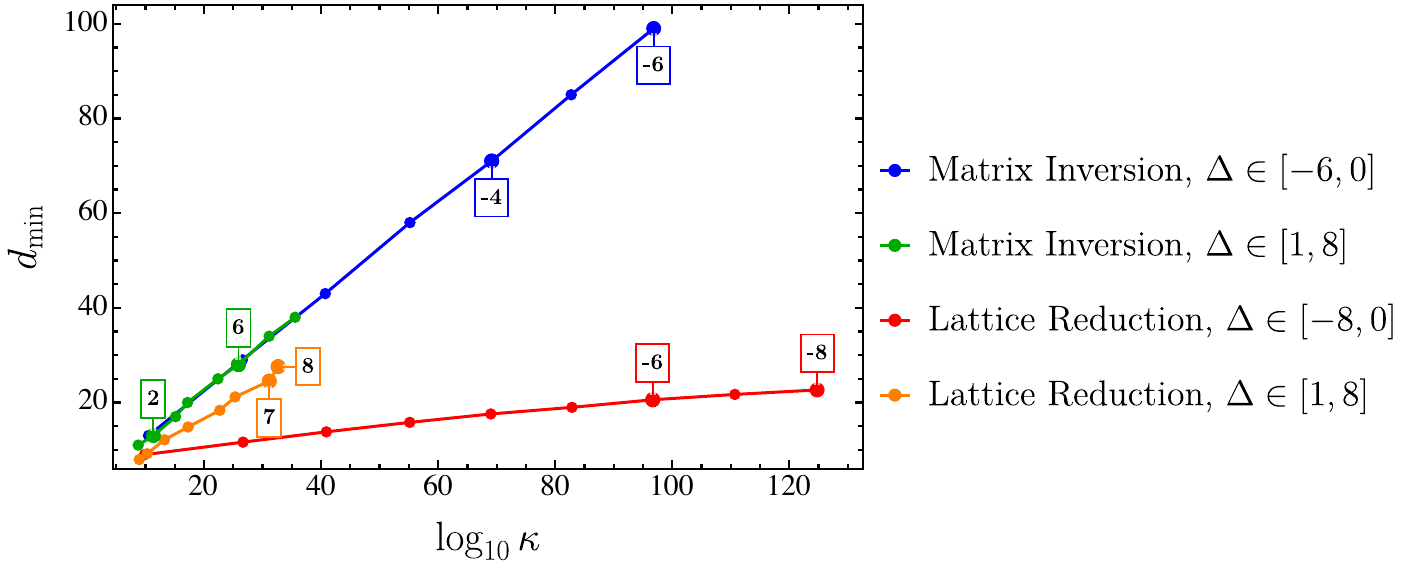}
    \caption{Minimum number of digits required for analytic regression using lattice reduction or linear matrix inversion as compared to condition number $\kappa$. The basis set are evaluated at 50 random points, sampled for $\frac{1}{2} - 10^{\Delta} \le x \le \frac{1}{2} + 10^{\Delta}$. The boxed labels on the plot indicate the value of $\Delta$ at each point. We see that for lattice reduction, $\log_{10} \kappa$ has a nearly linear relationship with the number of digits required, with the slope changing sign and magnitude if the points are close ($\Delta <0$) or far ($\Delta > 0$). With matrix inversion, the relationship is even more linear and moreover the magnitude of the slope does not change. The correlation coefficient of all four curves shown is $> 0.99$.
}
    \label{fig:kappa}
\end{figure}

The increase in precision needed when the function values among the points are too similar or too different is fairly well captured by the condition number.
The condition number $\kappa$ was mentioned in Eq.~\eqref{eq:condnum}. Its definition depends on the norm used. The standard $L^2$ norm of a matrix $A$ is defined as the maximum length of $A \cdot v$, where $v$ is any vector with unit length.
This matrix $L^2$ norm is alternatively called the {\textbf{maximum singular value}}, $\sigma_{\text{max}} = \|A\|$. The shortest value of $\|A\cdot v\|$ is called the {\bf minimum singular value}.
The {\textbf{condition number}} is the ratio of these
\begin{equation}
    \kappa (A)={\frac {\sigma _{\text{max}}(A)}{\sigma _{\text{min}}(A)}} \,.
\end{equation}
In Fig.~\ref{fig:kappa} we show the minimum number of digits required for different sampling ranges, plotted against the condition number. We see that indeed the condition number does correlate strongly with the minimum number of digits. 

In Fig.~\ref{fig:kappa}, we see that the condition number is exponentially large, reflecting the fact that our matrix is not random (which would lead to condition numbers of order one) but determined by evaluating smooth functions. Similarly exponentially large condition numbers can be seen with Vandermonde matrices which have entries determined by polynomials~\cite{Gautschi1974}. Fig.~\ref{fig:kappa} also shows the minimum digits needed if we use matrix inversion (which requires $p=n$). With matrix inversion, the condition number has complete predictive power of the number of digits.  The relevance of the condition number for lattice reduction is due to lattice reduction algorithms involving row-reduction similar to matrix-inversion algorithms.

\begin{figure}
    \centering
    \includegraphics[width=0.6\linewidth]{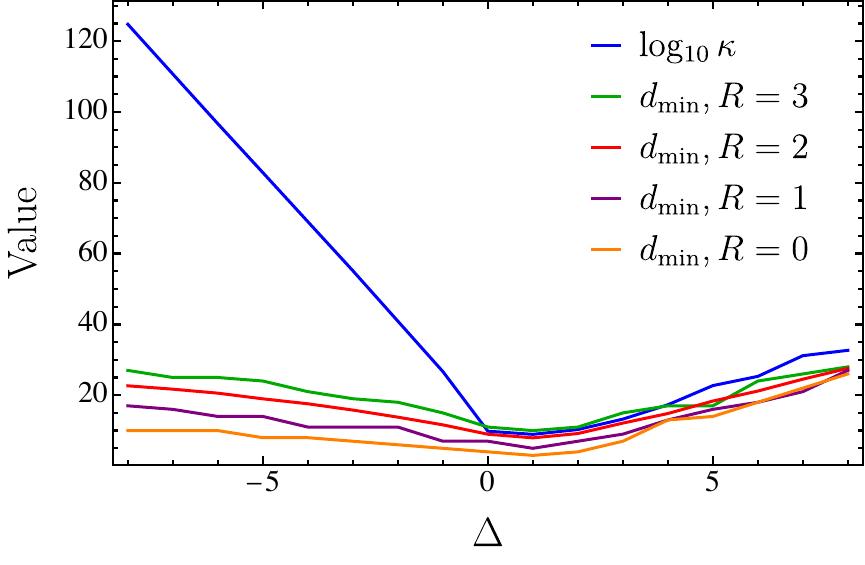}
    \caption{Value of $\log_{10}\kappa$, where $\kappa$ is the condition number, and the minimum number of digits needed $d_0$ using 50 points sampled within $\frac{1}{2} - 10^{-\Delta}\le x \le \frac{1}{2} + 10^{-\Delta}$. The same functions and conventions are used as in Figs.~\ref{fig:pxd} and~\ref{fig:plateau}.
    The coefficients of the true function are varied in the range $[-10^R,10^R]$. By construction, there is no dependence of $\kappa$ on $R$, since the 
    condition number is computed using only the basis functions and not  the true function.}
   
    \label{fig:comparison_delta}
\end{figure}

In order to minimize the number of digits needed, the simplest thing to do is to vary the sampling region. In Fig.~\ref{fig:comparison_delta} 
we  fix the number of functions to $n=15$ and number of points to $p=50$, and examine how $\dmin$ and $\kappa$ depend on the sampling region as set by $\Delta$ as above. 
We see that $\dmin$ and $\log_{10}\kappa$, to which it is strongly correlated, decrease linearly until $\Delta \approx 0$ before increasing linearly. 
For this figure, we compute the condition number using only the basis functions, not the true function $f$ or the coefficients $c_j$. Thus the condition number by construction cannot depend on $R$. The minimum number of digits does depend on $R$. However, for any choice of $R$, the optimum range of sampling values seems to still correspond to $\Delta \approx 1$. Such a conclusion of course depends on what the basis functions are (the optimum range for $\cB(\lambda x)$ for example will be scaled by $\lambda$ relative to the optimum range for $\cB(x)$). However, a useful rule of thumb from this analysis is that one should not take points too close together or too far apart.

\subsection{Optimizing point selection}
We have seen that lattice reduction allows for a tradeoff between digits of precision $d$ and number of points $p$, but that for a large number of points randomly selected in an interval, the precision requirement cannot be reduced below a floor $d_0$.
The saturation occurs because with too few digits the matrix of basis-function values becomes ill-conditioned (as characterized by the condition number). 
A rule-of-thumb from these observations is to avoid regions where the functions evaluated at the points are too similar or too distant. Similarly, one should also avoid picking points close to singularities of the basis functions, since those points will lead to large differences in basis function values. 
Despite these general observations, one might imagine it possible to reduce $\dmin$ below $d_0$ by judiciously picking points rather than randomly selecting them in an interval. 

To explore whether the number of digits required depends on the points, it is too computationally expensive to attempt gradient descent on the result of lattice reduction. So instead, we leverage the  strong correlation between $\dmin$ and condition number, which is computable from the matrix of basis functions alone. By replacing the full problem of lattice reduction with a simpler problem of computing the condition number we can in principle select the points ahead of time, using the basis functions alone, without the computationally expensive step of evaluating the true function. Because basis functions, particularly classical or multiple polylogarithms, can often be evaluated almost instantly, one can even search for optimal points.

\begin{figure}[t!]
    \centering
    \includegraphics[width=0.48\linewidth]{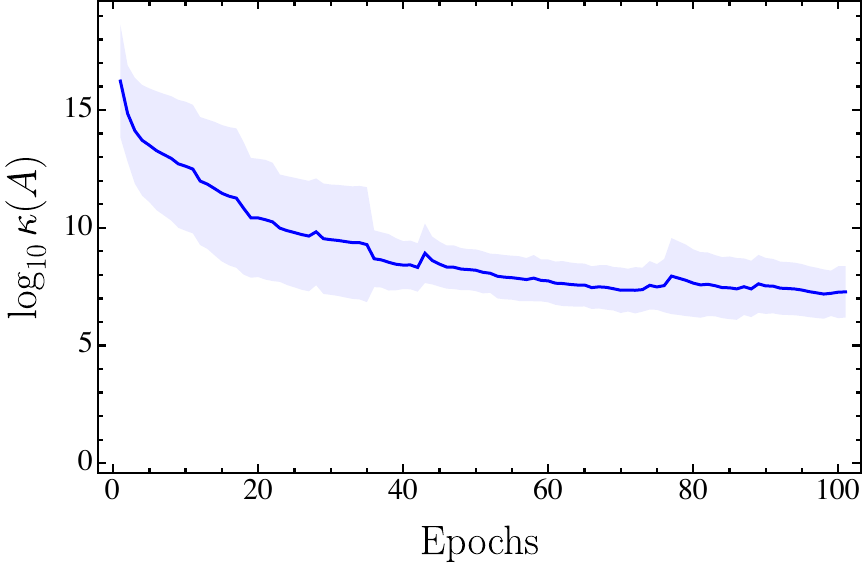}
    \includegraphics[width=0.48\linewidth]{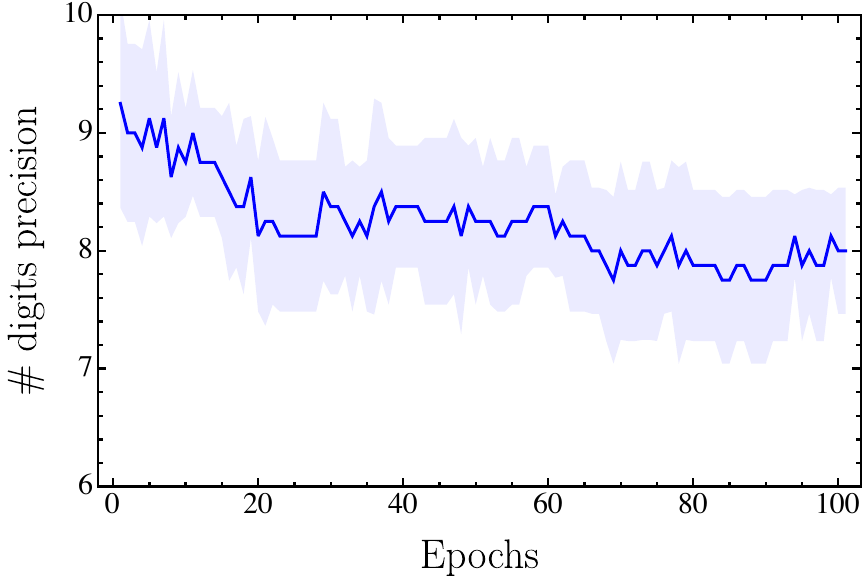}
    \caption{Gradient descent of lattice reduction for $n=15$, $p=15$. Minimizing the condition number by gradient descent allows us to reduce the number of digits by 1-2. }
    \label{fig:GDtest}
\end{figure}

To be concrete, we take our basis of $n=15$ GPLs as in Section~\ref{sec:p_d_scaling} with $p=15$ real points. Each set of 15 points leads to a matrix $\cB_i(x_j)$ and associated condition number $\kappa(\{x_j\})$. We then use gradient descent 
to minimize $\kappa$. In Fig.~\ref{fig:GDtest}, we show the results averaged over eight trials. We see that in fact the condition number does depend on the location of the points, and $\log_{10} \kappa$ can be lowered by about a factor of 2 this way. The corresponding $\dmin$ is reduced by 1-2 on average. For the most promising trial this point optimization takes us from requiring 10 digits of precision to requiring only 7. Such an approach could be critical if there is a very steep computational cost for improving the number of digits available for some particular application.

\subsection{Timing}\label{sec:timing_scaling}
In this section, we examine the computational cost of the various steps in analytic regression. We can compare the timing of naive  inversion, which requires the same number of points as functions ($p=n$) and PSLQ, which can only work with a single point ($p=1$), to lattice reduction, which can work in principle with any $p>0$. 
For this analysis, we consider the analytic regression of the outer-mass double-box integral in Section~\ref{sec:dbox} using a   set of GPLs in $w,z$ variables.

\begin{figure}
    \centering
    \includegraphics[width=0.7\linewidth]{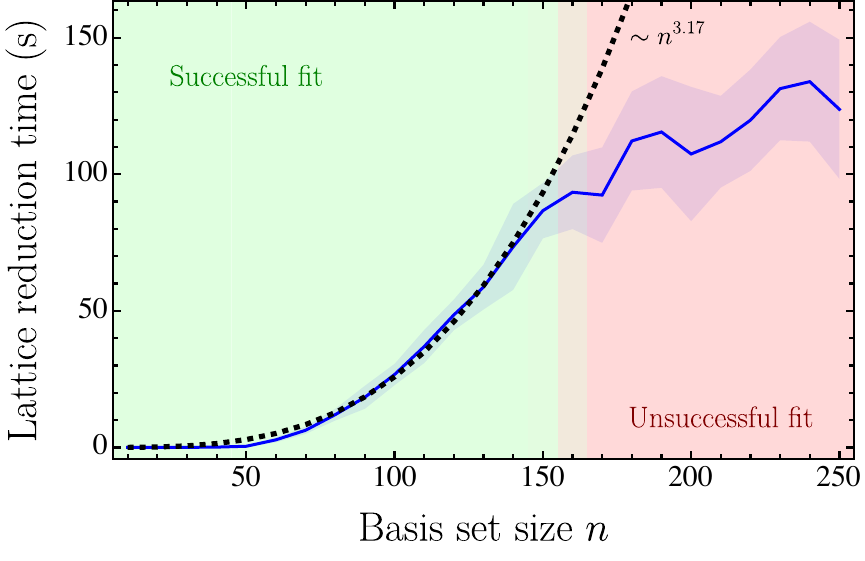}
    \caption{The blue curve shows the time required to perform the lattice reduction when fitting a subset of $n$ coefficients of the outer-mass double-box integral on $p=50$ points with $d=15$ digits of precision. The black dashed line a power-law fit, $t = n^{3.17}$. 
    The green shading indicates the region where lattice reduction successfully recovers all of the coefficients, while the red shaded region indicates an unsuccessful fit. The shaded blue band reflects the uncertainty over 12 trials with different basis vectors.}
    \label{fig:timings_lll1}
\end{figure}

First, we take $p=50$ random evaluation points in the  $(w, z) \in [0,1]$ range and look at the time taken by lattice reduction using 15 digits of precision as we vary the number of functions $n$. Fig.~\ref{fig:timings_lll1} shows that when lattice reduction successfully finds the appropriate coefficients, the scaling behavior is a regular power law ($t \sim n^{3.17}$ in this case). However, when two many basis functions are used for this precision ($n \gtrsim 150$) lattice reduction fails and the power law behavior breaks down. Although the timing in the $n \gtrsim 150$ region is below the anticipated power-law scaling, the actual practical computational complexity is in fact underestimated since lattice reduction does not successfully find the coefficients. Thus the timing in the red region of Fig.~\ref{fig:timings_lll1} is not so important.

\begin{figure}[t]
     \centering
     \hfill
         \includegraphics[width=0.48\textwidth]{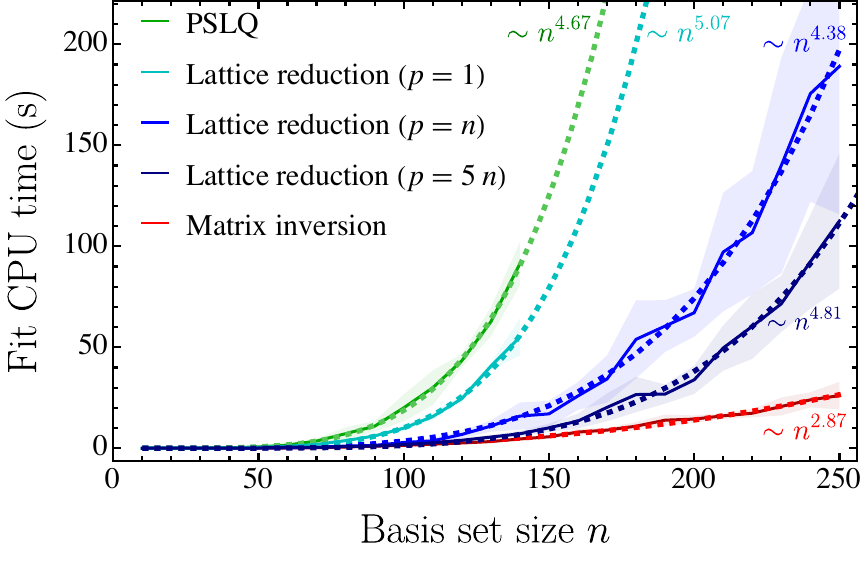}
     \hfill
         \includegraphics[width=0.48\textwidth]{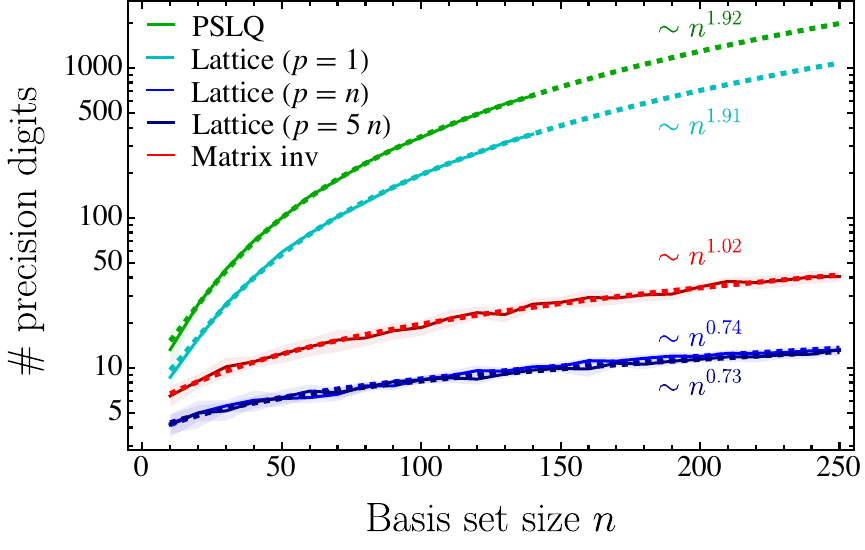}
     \hfill
     \caption{(Left panel) Time  required to perform either PSLQ (green), lattice reduction (blue) or matrix inversion (red). Matrix inversion requires the same number of evaluation points as basis functions ($p=n$). For PSLQ we use a single evaluation point ($p=1$). For lattice reduction we use either $p=1$, $p=n$ or $p=5n$. For all fits we require as many digits of precision as needed to have a successful fit. (Right panel) We indicate the number of digits of precision required/used, $d_\text{min}(n)$.}
              \label{fig:timings_all}
\end{figure}

To focus only on timing for examples where lattice reduction is successful, we first determine $\dmin(n)$, the minimal number of precision digits required to have a valid fit with $n$ basis functions, and then perform the reduction with $\dmin(n)$ precision. We also vary the number of evaluation points $p$, still taken in the $(w, z) \in [0,1]$ range. We consider the special cases  $p=n$ for an apples-to-apples comparison with the matrix-inversion approach and $p=1$ for a comparison to PSLQ, and  $p=5n$ to illustrate the domain where we have access to a large number of numerical samples. The resulting timings are displayed in the left panel of Fig.~\ref{fig:timings_all}. For matrix inversion we deem the regression to be successful if the result matches up with the actual double box coefficients with an error smaller than $10^{-2}$. We also include the timings associated with PSLQ, which can only make use of a single evaluation point, which can then be compared to the $p=1$  curve of lattice reduction. For each curve we also report the values of $\dmin(n)$ used in the right panel of Fig.~\ref{fig:timings_all}.

\begin{figure}[t]
     \centering
     \hfill
         \includegraphics[width=0.50\textwidth]{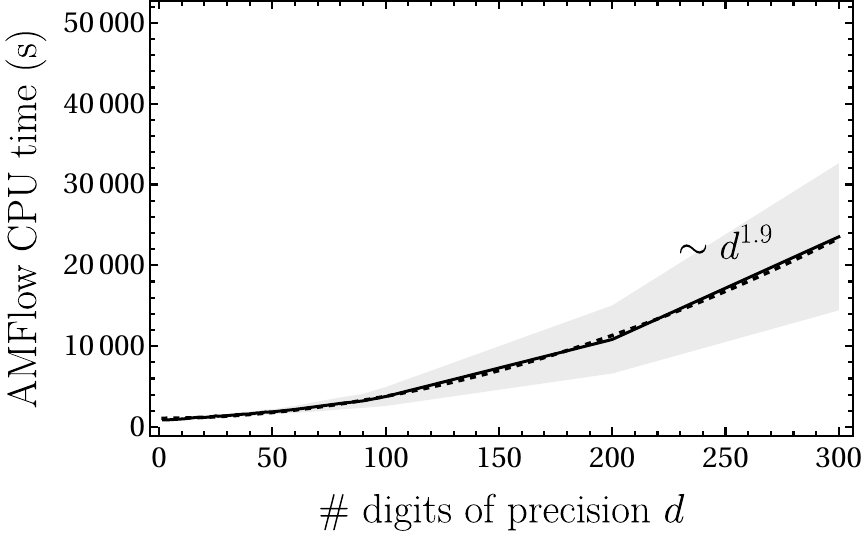}
     \hfill
         \includegraphics[width=0.47\textwidth]{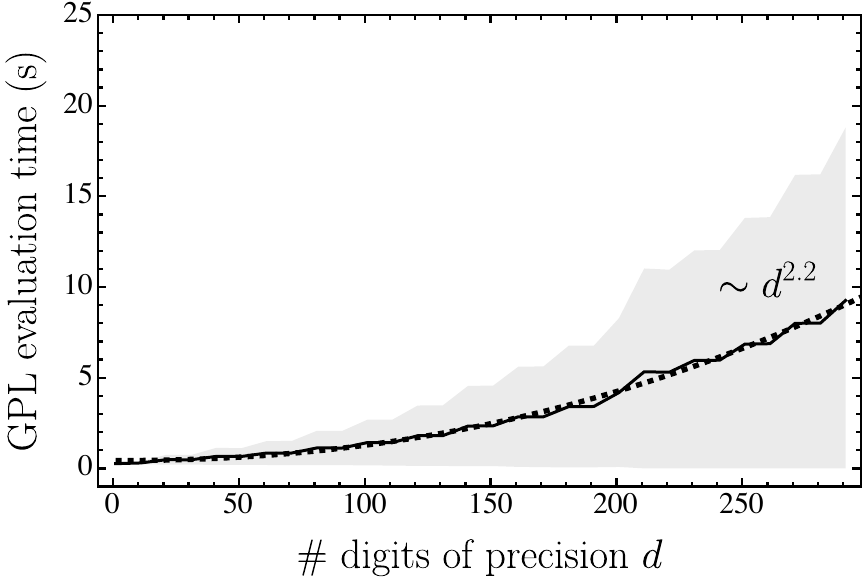}
     \hfill
     \caption{(Left panel) The time required to evaluate the double box integral $I_5$ of Eq.~\eqref{fig:twoloop_dbox} as a function of the number $d$ of digits of precision requested. The timing is given for \textsc{AMFlow}  with \textsc{Kira 3.0} used to perform the IBP reduction. (Right panel) The time for sampling the exact result of $I_5$ from its GPL expression using \textsc{Ginsh}. In both panels the shaded band represents the standard deviation in the sampling time across a set of 25 different kinematic points.
     }
    \label{fig:timings_entries}
\end{figure}

Fig.~\ref{fig:timings_all} also exhibits the precision plateau discussed in Section~\ref{sec:p_d_scaling}: as $p\geq n$ the precision required $d_\text{min}(n)$ is the same across lattice reduction runs. However, as can be observed from the left panel, the reduction time actually decreases with the number of evaluation points even though $d$ is fixed. However, accessing more sampling points requires more numerical evaluations which is not included in the lattice reduction timing. These costs are made more precise in the Fig.~\ref{fig:timings_entries}, where we show the evaluation times of respectively \textsc{AMFlow} and \textsc{Ginsh} at varying precision requirements. 
Both show power roughly quadratic scaling behavior ($t\sim d^2$). The timing for \textsc{AMFlow} includes IBP reduction to master integrals, IBP reduction to form DE with respect to the auxiliary mass $\eta$, and solving the DE.
For GPLs, the time cost for their evaluation through \textsc{Ginsh} is reasonably fast at low precision. 

In summary, the total computational cost for performing a reduction using $p$ points for $n$ basis functions in nanoseconds/CPU scales as  (very approximately)
\begin{equation}\label{eq:timing}
    t(p, n)/\text{ns} \approx  \underbrace{ 10^9 \cdot p  \cdot    d_\text{min}^2(n)}_{\textsc{AMFlow}} + \underbrace{10^4 \cdot n\cdot p \cdot d_\text{min}^{2}(n)}_{\textsc{Ginsh}} + \underbrace{\, p \cdot n^{4}}_{\text{fitting}} \,,
\end{equation}
where $\dmin(n)$ is as in Eq.~\eqref{numeric}. 
 The constants and the exponents are approximate and depend on the number of sampling points and the reduction algorithm used. To find the optimal number of sampling points $p$ at fixed $n$ one needs to balance these individual components  and account for the maximal precision attainable. 
 For a very large number of basis functions, both PSLQ and lattice reduction with $p=1$ have time costs dominated by both the \textsc{Ginsh} evaluations and the reduction, for a total time scaling close to $\sim n^{5}$. In contrast, the cost for lattice reduction with $p\geq n$ is solely driven by the reduction, with a scaling around $\sim n^{4.5}$. Matrix inversion, if accurate, would be preferred to these methods, since the time cost is dominated by the (parallelizable) $\textsc{Ginsh}$ evaluation, going as $n^{4.2}$. Unfortunately, matrix inversion is not guaranteed to return rational coefficients and is thus less trustworthy. Therefore, lattice reduction with $p\geq n$ emerges as the most robust method which has both a lower time complexity, and smaller precision requirements.

 Finally, we note that the evaluation of different kinematic points through \textsc{AMFlow} or \textsc{Ginsh} can easily be parallelized: the factors of $p$ and $np$ in Eq.~(\ref{eq:timing}) just represent the total number of evaluations necessary. The analytic regression techniques (matrix inversion, lattice reduction and PLSQ) are not so easy to parallelize although algorithmic improvements are possible.

\subsection{Summary of practical lessons}
We have explored in this section tradeoffs among different analytic regression methods and the minimum number of digits required $\dmin$ and number of sampling points $p$. The basic behavior we found for lattice reduction is that $\dmin \sim \frac{1}{p}+ d_0$: one can generally use fewer digits with more points, but there is a saturation effect and a minimum number of digits $d_0$ needed for even a large number of points. We found that the number of digits needed is determined largely by the condition number of the matrix of basis functions $\cB_i(p_j)$. To minimize the number of digits one should take the points not too close and not too far. The number of digits required is reduced moderately by trying to optimize point selection, but this may not be worth the effort in practice.

We have also studied the timings of different reduction algorithms: lattice reduction, matrix inversion, and PSLQ. In the low precision domain, lattice reduction with $p\geq n $ emerges as the best reduction algorithm. If high precision numerical samples are readily available, matrix inversion with rounding has the best formal scaling behavior, but is not recommended since it may fail (see Section~\ref{sec:mi}). PSLQ is strong for fitting numbers, but it cannot leverage the ability to sample functions at many points and therefore requires enormous precision so it is not a viable option. Thus lattice reduction emerges as the preferred algorithm. 

\section{Conclusions}
\label{sec:conclusion}
Recent years have witnessed an explosion in our ability to compute Feynman integrals both numerically and analytically. On the numeric side, public computer programs such as  \textsc{AMFlow} can compute extremely complicated Feynman integrals to dozens if not hundreds of digits of precision. On the analytic side, sophisticated mathematical techniques from algebraic geometry and elsewhere have manged to 
characterize well the space of functions where certain Feynman integrals must live. This paper attempts to bridge these two advancements by allowing for the regression of the exact analytic form Feynman integrals from high-precision numeric sampling.

The paper presents a number of end-to-end analytic regression examples, such as 3-loop triangle diagrams or the 2-loop outer mass double box. We compute these integrals exactly --  the full analytic form is obtained -- by combining a number of public codes for evaluating the integrals and basis functions with our analytic regression method. 

We consider three approaches to analytic regression: matrix inversion, lattice reduction, and PSLQ. Matrix inversion is the most straightforward: one samples the $n$ functions and $p=n$ points and solves the equations for the coefficients. This method requires rounding, which is imprecise, and suffers from a rapid loss of precision requiring large number of digits. PSLQ uses just a single point which requires tremendous precision (thousands of digits). Lattice reduction is a compromise: the number of points $p$ sampled can be small $p<n$ or large $p>n$ and it appropriately leverages the expectation that the coefficients are often order 1 rational numbers. 

We have also investigated crucial aspects such as the scaling behavior of lattice reduction, precision requirements, and the interplay between the number of evaluation points and numerical accuracy. We found that while increasing evaluation points substantially reduces precision requirements initially, there is a saturation point where precision improvements plateaus, influenced by the condition number and the distribution of sampled points. For integrals at the boundary of computational feasibility, one could attempt to judiciously select sampling points to optimize both accuracy and computational efficiency.

Looking forward, several directions appear promising for extending and refining this analytic regression methodology. Algorithmic improvements to lattice reduction techniques—such as parallelization or optimized variants—may allow for handling significantly larger basis sets and more complex integrals. Generalizing this approach beyond GPLs to other classes of special functions, such as elliptic polylogarithms, could further expand its applicability within theoretical physics and beyond. Additionally, one can imagine performing analytic regression even if the basis is not finite but instead constructable using some predicates, such as composition of polylogarithms and rational functions. In that way one could determine the alphabet and the final function at the same time. 

Finally, analytic regression not only provides practical tools for tackling challenging integrals encountered in high-energy physics but also holds potential for broader applications across various fields where exact analytic reconstruction from numerical data is sought. For example, one could envision applications in condensed matter physics, such as regressing a partition function based on high-precision numerical computation, or number theory, such as recovering relationships among multiple zeta values or modular forms.

\acknowledgments

MDS and AD would like to thank Michael Douglas and Siddharth Mishra-Sharma for valuable conversations at the start of this project. We also thank Miguel Correia, Johannes Henn, Xin Guan, Xiao Liu, and Tong-Zhi Yang for useful discussions.
The computations in this paper were run on the FASRC Cannon cluster supported by the FAS Division of Science Research Computing Group at Harvard University. AD and MDS are supported in part by the National Science Foundation under Cooperative Agreement PHY-2019786 (The NSF AI Institute for Artificial Intelligence and Fundamental Interactions). RH, MDS and XYZ are supported in part by the DOE Grant DE-SC0013607.

\appendix
\section{Lattice reduction }\label{app:lll_details}
In this appendix, we provide a brief introduction to  lattice reduction.  We refer to Refs.~\cite{Nguyen:2009b,Simon:2010} for a more detailed exposition.
Lattice reduction has many practical applications, for instance for factorizing polynomials \cite{Lenstra:1982}, the shortest vector problem \cite{Fincke:1985} or cryptanalysis \cite{Nguyen:2001}.
  Lattice reduction is an approach to the shortest distance problem, which asks for, given a lattice basis, the shortest vector contained in this lattice. Qualitatively, the idea is to start with a set of basis vectors that span a given lattice and to output an equivalent set of reduced basis vectors, which are shorter and almost orthogonal. The Lenstra–Lenstra–Lov\'asz (LLL) algorithm \cite{Lenstra:1982} is the first of these algorithms, proposed in 1982.

\subsection{LLL}

Given a set of $n$ linearly independent basis vectors $\mathcal{B}\equiv\{\vec b_1,\cdots \vec b_n\}$ in $\mathbb{R}^n$, a $n$-dimension lattice is defined as
\begin{equation}
    \mathcal{L}=\left\{\sum_{i=1}^n x_i \vec{b}_i \,|\, x_i \in \mathbb{Z}\right\}\,.
\end{equation}
The basis $\mathcal{B}$ is not unique and the LLL algorithm aims at searching for the shortest and most orthogonal basis. The reduction takes place by using a set of transformations that change the basis but preserve the lattice. Start with the original basis $\mathcal{B}=\{\vec b_1,\cdots \vec b_n\}$, we can first derive the orthogonal basis $\mathcal{B}^\prime=\{\vec b_1^\prime,\cdots \vec b_n^\prime\}$ through the Gram-Schmidt orthogonalization. The explicit form is 
\begin{align}
    \vec b_1^\prime &= \vec b_1\,,\notag\\
    \vec b_2^\prime &= \vec b_2 - \mu_{2,1} \vec b_1\,,\notag\\
    \vec b_3^\prime &= \vec b_3 - \mu_{3,1} \vec b_1- \mu_{3,2} \vec b_2\,,\notag\\
    \cdots\notag\\
    \vec b_n^\prime & = \vec b_n - \sum_{i=1}^{n-1}\mu_{n,i}\vec b_i\,,
\end{align}
where we introduce the projection coefficient $\mu_{i,j}$
\begin{equation}
    \mu_{i,j}=\frac{\langle \vec b_i, \vec b_j^\prime\rangle}{\langle \vec b_j^\prime, \vec b_j^\prime\rangle}= \frac{\vec b_i \cdot \vec b_j^\prime}{\vec b_j^\prime \cdot \vec b_j^\prime}\,.
\end{equation}
$\mathcal{B}$ is called a LLL-reduced basis if and only if it satisfies the two conditions:
\begin{itemize}
    \item If $i>j$, then $|\mu_{i,j}|\leq \frac{1}{2}$
    \item If $i<n$, then $\left(\delta-\mu_{i+1,i}^2\right)\|\vec b_i^\prime\|^2<\|\vec b_{i+1}^\prime\|^2$
\end{itemize}
where $\delta$ is a parameter of LLL and typically $\frac{1}{4}<\delta<1$.
The first inequality is usually referred to as the projection condition and the second one is the Lov\'asz condition.  In Alg.~\ref{alg:lll}, we present the pseudocode for the LLL algorithm, where the size reduction aims at approaching the projection condition and the swapping is for the Lov\'asz condition.

\begin{algorithm}[!htbp]
  \caption{LLL algorithm}\label{alg:lll}
  \begin{algorithmic}[1]
    \Procedure{LLL}{$\mathcal{B} = [b_1,\dots,b_n],\,\delta$}
      \State  Compute Gram–Schmidt orthogonalization of $B$: 
        \For{$i = 1,\dots,n$}
          \State $\displaystyle b_i^\prime \gets b_i - \sum_{j=1}^{i-1} \mu_{i,j}\, b_j^\prime$
          \State where $\displaystyle \mu_{i,j} = \frac{\langle b_i, b_j^\prime\rangle}{\langle  b_j^\prime,b_j^\prime\rangle}$
        \EndFor
      \State $k \gets 2$
      \While{$k \le n$}
        \Comment{Size‐reduce $b_k$ against $b_{k-1},\dots,b_1$}
        \For{$j = k-1,\dots,1$}
          \If{$|\mu_{k,j}| > \tfrac12$}
            \State $b_k \gets b_k - \mathrm{round}(\mu_{k,j})\,b_j$
            \State Recompute $\mu_{k,\ell}$ for $\ell<k$, but do not recompute Gram–Schmidt
          \EndIf
        \EndFor
        \Comment{Lov\'asz condition}
        \If{$\| b_k^\prime\|^2 < (\delta - \mu_{k,k-1}^2)\,\| b_{k-1}^\prime\|^2$}
          \State Swap $b_k \leftrightarrow b_{k-1}$
          \State Recompute Gram–Schmidt for indices $k-1,k$
          \State $k \gets \max(k-1,2)$
        \Else
          \State $k \gets k + 1$
        \EndIf
      \EndWhile
      \State \Return {$\mathcal{B}$ (LLL‐reduced basis)}
    \EndProcedure
  \end{algorithmic}
\end{algorithm}

Although the LLL algorithm is originally designed for an integer lattice basis, it can also handle complex numbers. This requires modifying the dot product $\langle b_i, b_j^\prime\rangle= b_i \cdot (b_j^\prime)^\star$ and the rounding in the size reduction step $\text{round}(\mu_{k,l})=\text{round}(\text{Re}\,\mu_{k,l})+ i \, \text{round}(\text{Im}\,\mu_{k,l})$. This feature has been included in the \textsc{Mathematica} built-in command \textsc{LatticeReduce}.

Algorithmic improvements to the LLL algorithm~\cite{ Kaltofen:1983,  Schnorr:1988, Gama:2006, Nguyen:2009,Dabral:2024} aim to reduce  runtime and/or output quality. A popular variant  is the L${}^2$ algorithm~\cite{Nguyen:2009} which is default algorithm in \textsc{Mathematica} and \textsc{Fplll}.

 \subsection{A simple example of LLL reduction}\label{sec:app_simple_example_lll} 
We consider the function $f(x) = G(0,1;x)- G(1,-1;x)$ given as a linear combination of two weight 2 GPLs which depend on a single argument $x$. For the basis functions  we retain the set $\mathbf{\mathcal{B}}(x)=\left\{G(1,0;x),G(0,1;x), G(0,-1;x), G(1,-1;x) \right\}$, containing 4 different functions. We evaluate $f$ and $\mathbf{\mathcal{B}}$ at two evaluation points, $x_1=4/10$ and $x_2=9/10$. Using $3$ digits of precision, the matrix of Eq.~(\ref{eq:mat_lll}) reads\footnote{In our example we have used $d=3$, $n=4$ and $p=2$. The largest entry is $\mathcal{B}_1(x_1)=1.54...$, giving $s=d-v_\text{max}=3-1=2$, so that before rounding we multiply all entries by $10^2$.}  
\begin{equation}
    M=\begin{pmatrix}
        \mathbf{v}_1\\
        \mathbf{v}_2\\
        \mathbf{v}_3\\
        \mathbf{v}_4\\
        \mathbf{v}_5
    \end{pmatrix}=\left(
\begin{array}{ccccccc}
 -35 & -24 & 1 & 0 & 0 & 0 & 0 \\
 92 & 154 & 0 & 1 & 0 & 0 & 0 \\
 -45 & -129 & 0 & 0 & 1 & 0 & 0 \\
 36 & 75 & 0 & 0 & 0 & 1 & 0 \\
 -10 & -106 & 0 & 0 & 0 & 0 & 1 \\
\end{array}
\right) \,,
\end{equation}
which gets LLL reduced to 
\begin{equation}
    \tilde{M} =\begin{pmatrix}
        \mathbf{u}_1\\
        \mathbf{u}_2\\
        \mathbf{u}_3\\
        \mathbf{u}_4\\
        \mathbf{u}_5
    \end{pmatrix}= \left(
\begin{array}{ccccccc}
 0 & -1 & 1 & 0 & -1 & 0 & 1 \\
 4 & -2 & 2 & 1 & 2 & 2 & 0 \\
4 & 4 & -2 & -1 & -4 & -4 & 1 \\
 0 & -1 & 4 & 6 & 4 & -7 & -2 \\
1 & -10 & -5 & 1 & 2 & -6 & -4 \\
\end{array}
\right) \,.
\end{equation}
We can verify explicitly, by computing the Gram–Schmidt process of $\tilde{M}$, that the new basis is size-reduced, in that $|\mu_{i_j}|<1/2$ for $ 1\leq j < i \leq 5$. It also obeys the Lov\'asz  condition, $\| b_k^\prime\|^2 < (\delta - \mu_{k,k-1}^2)\,\| b_{k-1}^\prime\|^2$ for $2\leq k\leq 5$. Although we did not explicitly optimize for the size reduction of the $\mathbf{u}_i$, we can verify that the new basis vectors are all reduced (and ordered) with respect to their Euclidean norm. This ordering is not guaranteed, but tends to be the case for non-pathological lattices where the Gram–Schmidt coefficients are much smaller than their worst-case bound of $1/2$. We also note that the final form of $\tilde{M}$ depends on the ordering of the input basis $\mathbf{v}_i$, since the Gram–Schmidt process depends on the ordering of the original basis vectors. In practice this also has minimal impact for our results, since we are mostly concerned with the shortest vector of $\tilde{M}$. Numerically we observed these effects to change the precision requirements of a given fit by a couple of digits of precision at most.

The first row of $\tilde{M}$, $\mathbf{u}_1$, can by definition be written in terms of the original basis. Focusing on the last 5 entries we can immediately read off $\mathbf{u}_1 = \mathbf{v}_1 - \mathbf{v}_3 + \mathbf{v}_5$, which in turn implies 
\begin{align}
    0 =&  f(x_1) - G(0,1;x_1) + G(1,-1;x_1) \\
    -1 =&  10^2 \left[f(x_2) - G(0,1;x_2) + G(1,-1;x_2) \right] \label{eq:f2_lll_red} \,, 
\end{align}
from which we can extract the relation for $f(x)$. The fact that Eq.~(\ref{eq:f2_lll_red}) does not quite evaluate to 0 is not an issue, but rather a reflection of the fact that our LLL reduction was only taken with finite precision.

\section{The symbol formalism}\label{sec:symbols}

In this appendix, we review the Generalized polylogarithms (GPLs) and the symbol formalism. The GPL is defined iteratively by 
\begin{equation}\label{eq:gpl_def}
G(a_1,\cdots a_n; x)\equiv\int_0^x \frac{dt}{t-a_1} G(a_2,\cdots a_n; t)\,,
\end{equation}
with
\begin{equation}
G(;x)\equiv1,\quad G(\vec 0_n;x)\equiv\frac{1}{n!}\ln^n (x)\,.
\end{equation}
It is conjectured that GPLs up to transcendentality-three can be expressed in terms of logarithms and classical polylogarithms $\Li_n(x)$ with $n\leq 3$ \cite{Duhr:2011zq}. For transcendentality-four GPLs, one also needs the special function $\Li_{2,2}(x,y)$. The conversion can be done with public packages like \textsc{PolyLogTools} \cite{Duhr:2019tlz} or \textsc{Gtolrules} \cite{Frellesvig:2016ske}. There are also public packages for numerically evaluating the GPLs: \textsc{GiNac}~\cite{Bauer:2002, Vollinga:2004sn}, \textsc{FastGPL}~\cite{Wang:2021imw} and \textsc{handyG}~\cite{Naterop:2019xaf}.

Manipulating or simplifying GPLs/polylogarithmic expressions with identities is a non-trivial task in practice. In Ref.~\cite{Goncharov:2010jf}, the symbol formalism is introduced to simplify the notation and allows for imposing the identities in an easier way. First of all, notice that the GPL can also be written in terms of the dlog form
\begin{equation}
      G (a_1, \cdots, a_n ; x) =\int_0^x d \ln (x-a_1) \circ \cdots \circ d \ln (x-a_n) \,.
\end{equation}
Then the symbol is defined to be
\begin{equation}
    \cS\left[\int_a^b d \ln R_1  \circ \cdots \circ d \ln R_n \right]
    =R_1 \otimes \cdots \otimes R_n
\end{equation}
so that
\begin{equation}\label{symbolpolylogs}
  \cS\big[G(a_1,\ldots,a_n;x)\big] = (x-a_1) \otimes \cdots \otimes (x-a_n)
\end{equation}
with the special cases
\begin{equation}\label{symbolpolylogs2}
  \cS\big[\Li_n (x)\big] = - (1 - x) \otimes \underbrace{x \otimes
  \cdots \otimes x}_{n - 1},\quad 
    \mathcal{S} \left[\frac{1}{n!}\ln^n x\right] = \underbrace{x \otimes
  \cdots \otimes x}_{n} \,.
\end{equation}
Note that the symbol acting on a complex number vanishes, e.g.  $\cS[c]=0,\,c\in\mathbb{C}$. In other words, symbol only captures the leading transcendentality of the GPLs. Therefore, when constructing the function basis, we need to add terms like $\zeta_n\times$GPL for completeness.

Symbol satisfies the following properties:
\begin{align}
\text{Product rule: } \quad & \cdots \otimes f(x)g(x) \otimes \cdots 
= \cdots \otimes f(x) \otimes \cdots + \cdots \otimes g(x) \otimes \cdots \notag\\
\text{Blind to constants: } \quad & \cdots \otimes cf(x) \otimes \cdots 
= \cdots \otimes f(x) \otimes \cdots \\
\text{Shuffle product: } \quad & \mathcal{S}(f \cdot g) = \mathcal{S}(f) \shuffle \mathcal{S}(g)\notag
\end{align}
and they have already encoded all the polylogarithmic identities. For example, symbol for the weight-2 five-term identity:
\begin{align}
    &\mathcal{S}\Big[\Li_2(x) + \Li_2(y) + \Li_2\left(\frac{1 - x}{1 - x y}\right)
+ \Li_2(1 - x y) + \Li_2\left(\frac{1 - y}{1 - x y}\right)\Big]\notag\\
=&\mathcal{S}\Big[\frac{\pi^2}{2} - \log(x) \log(1 - x) - \log(y) \log(1 - y)
- \log\left( \frac{1 - x}{1 - x y} \right) \log\left( \frac{1 - y}{1 - x y} \right)\Big]
\end{align}
These properties will enable us to simplify polylogarithmic expressions in a more straightforward way~\cite {Dersy:2022bym}. 

In our bootstrap program, after constructing all possible symbols up to a certain weight, we also need to integrate them back into GPLs. To do so, we need to ensure all symbol bases are integrable, which is determined by the {\it integrability conditions}. Given a symbol expression:
\begin{equation}
    S=\sum_{i_1,i_2,\cdots i_N}c_{i_1,i_2,\cdots i_N} a_{i_1}\otimes a_{i_2}\cdots \otimes a_{i_N}\,,
\end{equation}
the integrability condition states
\begin{equation}
    \sum_{i_1,i_2,\cdots i_N}c_{i_1,i_2,\cdots i_N} a_{i_1}\otimes a_{i_2}\cdots \otimes a_{i_{p-1}}\otimes a_{i_{p}}\cdots \otimes a_{i_N} \, d\log a_{i_{p-1}} \wedge d\log a_{i_p} =0\,,
\end{equation}
for any pair $\{i_{p-1},i_p\}$ with $1<p\leq N$. By solving these equations, we obtain the integrable symbol basis at each weight and construct the function basis as in Eq.~\eqref{eq:basis_tower}. 

A systematic construction of the symbol basis is as follows. First of all, all weight-one symbols can be integrated into logarithmic functions themselves. For weight-two, we compute these wedge products for any pair of the weight-2 symbols, evaluate them with some random numbers sampled and search for the null space. This amounts to giving all integrable weight-two symbols. Starting from weight-$n$ ($n\geq 3$), we take the integrable symbol basis $\{S^{(n-1)}_i\}$ at one weight lower, attach an additional letter $L_j\in \mathcal{A}$ from the alphabet before the symbol (or after the symbol, but not both): $\{L_j\otimes S^{(n-1)}_i\}$ or $\{S^{(n-1)}_i\otimes L_j\}$. Then we make an ansatz for weight-$n$ symbol: $S^{(n)}=\sum_{i,j}c_{i,j}L_j\otimes S^{(n-1)}_i$, with some unknown constants $c_{i,j}$ to fix, and apply the integrability conditions. This leads to a linear system on $c_{ij}$, and the solutions correspond to all possible integrable weight-$n$ symbols. This approach will allow us to iteratively construct the symbol basis. Notice that in order to reduce the size of the basis, we also apply lattice reduction at each weight during the construction. 

\section{Numerically evaluating iterated integrals}\label{sec:app_iterated_integrals}
In this appendix we review the numerical integration of the iterated integrals of the form of Eq.~(\ref{eq:iterated_int_symbol}) and highlight some of the key challenges. 

As a first example we consider the integration based on the symbol entries 
\begin{equation}\label{eq:symb_8,12,4,12}
    \mathcal{S}_{8,12,4,12} = \frac{\beta _v-1}{\beta _v+1}\otimes\frac{\beta _{uv}-\beta _u \beta _v}{\beta _{uv}+ \beta _u \beta _v}\otimes 1+v\otimes \frac{\beta _{uv}-\beta _u \beta _v}{\beta _{uv}+\beta _u \beta _v} \,,
\end{equation}
where the associated integral $\tilde{B}_{8,12,4,12}$ is well defined using the path $(u(\lambda), v(\lambda))=(u/\lambda, v/\lambda)$. The first symbol entry does not vanish nor blow up as $v\rightarrow \infty$, that is at $\lambda= 0$, and gives rise to a branch cut in the integral as opposed to a pole. Indeed we can check that the integrand term evaluates to $\partial_\lambda \log L_8(\lambda)= - \lambda^{-1/2} (v+\lambda)^{-1/2}$ for the corresponding entry. Such integrals pose no difficulty in their numerical evaluation. Moving instead to another less trivial example, we look at 
\begin{equation}\label{eq:symb_4,4,4,2}
    \mathcal{S}_{4,4,4,2}= 1+v \otimes 1+v \otimes 1+v \otimes v
\end{equation}
where now the first symbol entry blows up as $v\rightarrow \infty$ and is ill-defined for the path $(u(\lambda), v(\lambda))=(u/\lambda, v/\lambda)$ at the boundary $\lambda=0$. We can check that in the associated integral the corresponding integrand term goes as $\partial_\lambda \log L_4(\lambda) = - v \lambda^{-1} (v+\lambda)^{-1}$ , exhibiting a simple pole at $\lambda=0$. Therefore we cannot naively take the numerical integral for $\tilde{B}_{4,4,4,2}$. One way to get around this difficulty is to simply consider a different path, with different boundary conditions. For instance, if we instead parametrize our path as $(u(\lambda), v(\lambda))=(u \lambda, v \lambda)$, the first symbol entry is now finite at $\lambda = 0$. The integrand term doesn't have a simple pole and the numerical integral can be evaluated. Naturally this comes at a cost. The new path has a different boundary condition and it may be difficult to relate different integrals that have been evaluated with different boundary conditions.

As a final example, to connect with the existing literature on trailing zeros (see \cite{Walden:2020odh} for instance), we consider 
\begin{equation}    \label{eq:symb_2,4,4,2}
    \mathcal{S}_{2,4,4,2}= v \otimes 1+v \otimes 1+v \otimes v\,,
\end{equation}
where both paths $(u/\lambda, v/\lambda)$ and $(u \lambda, v \lambda)$ lead to a pole in the integrand of the associated integral at $\lambda =0$. To proceed we could naturally define another new path, but instead one can realize that the symbol entry of Eq.~(\ref{eq:symb_2,4,4,2}) corresponds to a particular GPL
\begin{equation}
    \mathcal{S}\left[G(0,-1,-1,0;v)\right] = \mathcal{S}_{2,4,4,2} \,.
\end{equation}
This allows us to utilize the shuffle relation (see \cite{Duhr:2011zq,Duhr:2012fh} for the general definition) 
\begin{equation}
    G(0,-1,-1,0;v)= G(0;v)\,G(0,-1,-1;v)-G(0,-1,0,-1;v)-2G(0,0,-1,-1;v) \,,
\end{equation}
leading to 
\begin{equation}
    \tilde{B}_{2,4,4,2} = (\log v) \tilde{B}_{4,4,2} - \tilde{B}_{4,2,4,2} - 2\tilde{B}_{4,4,2,2}  \,.
\end{equation}
Now each individual integral can be evaluated on the integration path $(u(\lambda), v(\lambda))=(u \lambda, v \lambda)$. It is however more difficult to systematically generalize this procedure in the multi-variable case we are considering where the symbols do not always correspond to a simple GPL. 

In the bootstrap example of Section~\ref{sec:dbox} the situation is slightly simpler, since we know that the final answer must vanish for $u,v \rightarrow \infty$. Therefore, one possible prescription for performing the numerical integration of symbols is to retain only linear combinations whose first letter does not lead to an ill-defined integral for the double-box boundary conditions. All letters of the alphabet of Eq.~(\ref{eq:alphabet_uv}) that involve a square root are not problematic and can appear in the first entry of the symbol. The letters linear in $u$ and $v$ require more care, and we find that the combinations
\begin{equation}
    \left\{\frac{u}{1+u},\frac{v}{1+v}, \frac{1+u}{1+u+v}, \frac{1+v}{1+u+v}, \frac{u+v}{1+u+v}\right\}
\end{equation}
lead to well behaved integrals, since the associated integrands do not have a pole in $\lambda$ for the path $(u(\lambda), v(\lambda))=(u/ \lambda, v/ \lambda)$. Therefore, we have a basis of 11 independent letters that can appear in the first entry of the symbol, purely from boundary conditions considerations. It turns out that all basis functions defined by imposing integrability, galois symmetry and physical branch cuts satisfy this property.

\bibliographystyle{JHEP}
\bibliography{biblio.bib}
\end{document}